\documentclass[preprint]{ptephy_v1}

\preprintnumber{XXXX-XXXX} 
\usepackage{graphicx}	
\usepackage{amsmath}	
\usepackage{amssymb}	
\usepackage{natbib}
\usepackage{hyperref}
\usepackage{subfigure}
\usepackage{romannum}
\usepackage{cancel}
\usepackage{ulem}

\newcommand{\diff}{{\text{d}}}

\usepackage{color}

\begin{document}

\title{
Time-Dependent Accretion Disks with Magnetically Driven Winds: Green’s Function Solutions
}



\author{Mageshwaran Tamilan}
\affil{Manipal Centre for Natural Sciences, Manipal Academy of Higher Education, Manipal 576104, India
\email{mageshwaran.tamilan@manipal.edu,~tmageshwaran2013@gmail.com}}

\begin{abstract}%

We present Green’s function solutions for a geometrically thin, one-dimensional Keplerian accretion disk that includes angular momentum extraction and mass loss due to magnetohydrodynamic (MHD) winds. The disk viscosity is assumed to vary radially as $\nu \propto r^{n}$. We derive solutions for three types of boundary conditions applied at the inner radius $r_{\rm in}$: (i) zero torque, (ii) zero mass accretion rate, and (iii) finite torque and finite accretion rate, and investigate the time evolution of a disk with an initial surface density represented by a Dirac-delta function. The mass accretion rate at the inner radius decays with time as $t^{-3/2}$ for $n = 1$ at late times in the absence of winds under the zero-torque condition, consistent with Lynden-Bell \& Pringle (1974), while the presence of winds leads to a steeper decay. All boundary conditions yield identical asymptotic time evolution for the accretion and wind mass-loss rates, though their radial profiles differ near $r_{\rm in}$. Applying our solutions to protoplanetary disks, we find that the disk follows distinct evolutionary tracks in the accretion rate–disk mass plane depending on $\psi$, a dimensionless parameter that regulates the strength of the vertical stress driving the wind, with the disk lifetime decreasing as $\psi$ increases due to enhanced wind-driven mass loss. The inner boundary condition influences the evolution for $\psi < 1$ but becomes negligible at higher $\psi$, indicating that strong magnetically driven winds dominate and limit mass inflow near the boundary. Our Green’s function solutions offer a general framework to study the long-term evolution of accretion disks with magnetically driven winds.
\end{abstract}

\subjectindex{E10, E14, E31, E34, E36}

\maketitle

%


%
\section{Introduction} 
\label{sec:intro}
%

The theory of astrophysical accretion disks has been applied to explain the emission properties of protoplanetary disks, active galactic nuclei, X-ray binaries, and tidal disruption events. When the gravitational potential is dominated by a central compact object and the viscous dissipation timescale is much longer than the orbital timescale, the accretion flow rapidly adjusts to the central potential and attains a quasi-Keplerian rotation profile \cite{2002apa..book.....F,2008bhad.book.....K}. On the longer viscous timescale, the flow evolves slowly and remains nearly axisymmetric at each radius. If radiative cooling is efficient, the accretion flow becomes geometrically thin. In the standard model of accretion disks, hydrodynamic eddy turbulence acts as a source of viscosity that redistributes angular momentum within the disk \cite{1973A&A....24..337S}. By introducing the $\alpha$-parameter to describe this turbulent viscosity, Shakura \&  Sunyaev (1973) \cite{1973A&A....24..337S} developed a steady-state solution for a geometrically thin, optically thick disk, in which the viscosity is expressed as a function of radius and surface density.

Lynden-Bell and Pringle (1974) \cite{1974MNRAS.168..603L} presented an analytical Green's function solution for the time evolution of an accretion disk, assuming the kinematic viscosity follows a power-law dependence on radius $r$. Their solution satisfies a boundary condition of either zero torque or zero mass flux at the origin ($r_{\rm in}=0$, where $r_{\rm in}$ denotes the inner edge of the disk). However, in actual astrophysical systems, the central object has a finite size, and the choice of inner boundary condition can significantly affect the disk's observable properties. In particular, the luminosity and variability timescales of black hole accretion flows are strongly influenced by the location of the innermost stable circular orbit (ISCO). Addressing this limitation, Tanaka (2011) \cite{2011MNRAS.410.1007T} extended the classical theory by obtaining exact Green’s function solutions for a disk with a finite inner radius ($r_{\rm in} > 0$), while still assuming a power-law dependence of viscosity on radius.
These solutions characterize a disk that evolves purely through accretion onto the central object, with no mass loss through winds.

The presence of magnetic fields in accretion disks can trigger significant instabilities due to the differential rotation of electrically conducting fluids. One such instability is the magnetorotational instability (MRI; \citealp{Velikhov1959,1991ApJ...376..214B,1998RvMP...70....1B}), which arises when weak magnetic fields interact with the disk's shearing motion. MRI drives magnetohydrodynamic (MHD) turbulence, resulting in stresses—both magnetic (Maxwell) and kinetic (Reynolds)—that transport angular momentum and allow matter to spiral inward. The mean flow behavior of MHD turbulence in an accretion disk aligns with the $\alpha$ prescription \cite{1999ApJ...521..650B}. Moreover, turbulence in magnetized disks can launch vertical outflows, the strength of which is closely tied to the magnetic field intensity in the disk \citep{2009ApJ...691L..49S,2014ApJ...784..121S,2013ApJ...765..149C,2019ApJ...872..149L}. As MRI develops, it can amplify initially weak magnetic fields and form coherent channel flows where magnetic pressure becomes comparable to the gas pressure. These flows eventually break down due to magnetic reconnection, producing episodic, turbulent-driven winds \citep{2009ApJ...691L..49S}. Additionally, if strong poloidal magnetic fields are present and rotate with the disk, they can centrifugally accelerate gas along magnetic field lines. When the angle between the poloidal field and the disk surface exceeds 60 degrees, this mechanism—often referred to as magnetocentrifugal acceleration—can drive sustained outflows \citep{1982MNRAS.199..883B}. A clear distinction between the magnetic pressure-driven and magnetocentrifugal processes is lacking and they are expected to work together cooperatively. These magnetically induced winds play a crucial role in regulating the disk's evolution by carrying away mass, angular momentum, and energy, thereby influencing its structure and emission characteristics.

Suzuki et al. (2016) \cite{2016A&A...596A..74S} developed a time-dependent, one-dimensional (1D) model of a geometrically thin accretion disk with magnetically driven winds, focusing on protoplanetary disks. Their formulation employed an extended $\alpha$-prescription to model magnetohydrodynamic (MHD) turbulence stresses in both radial and vertical directions. Building on this, Tamilan et al. (2024) \cite{2024ApJ...975...94T} performed numerical simulations of a similar 1D, time-dependent thin disk with magnetically driven winds, applied to TDEs. Tabone et al. (2022) \cite{2022MNRAS.512.2290T} derived self-similar solutions for such disks by assuming a sound speed profile of $c_s^2 \propto r^{-3/2+\gamma}$, where $\gamma$ is a constant power-law index, and a vertically time-independent sound speed structure. In the absence of winds, their solution recovers the late-time viscous disk evolution described by Lynden-Bell and Pringle (1974) \cite{1974MNRAS.168..603L}. More recently, Shadmehri and Khajenabi (2024) \cite{2024MNRAS.528.3294S} obtained an analytical solution for a time-dependent disk with magnetically driven winds, under the assumptions of a sound speed scaling as $c_s \propto r^{-1/2}$ and negligible radial turbulent viscosity, such that disk evolution and accretion are driven solely by vertical stress.

Protoplanetary disks are rotationally supported structures of gas and dust surrounding young, pre–main-sequence stars, with initial gas masses typically ranging from a few percent up to about 10\% of the stellar mass \cite{2011ARA&A..49...67W,2022A&A...657A..74F}. Observations from multi-object spectroscopy of nearby star-forming clusters spanning ages from 2 to 30 Myr show that the fraction of accreting stars declines from roughly 60\% at 2 Myr to about 2\% at 10 Myr, with no detectable accretion beyond this age \cite{2010A&A...510A..72F}. The long-term evolution of these disks is governed by the redistribution of angular momentum through viscous and magnetically mediated stresses, together with mass loss through disk winds. Recent analytical models, such as the self-similar solution developed by Tabone et al. (2022)\cite{2022MNRAS.512.2290T}, have provided valuable insights into these coupled processes. In this work, we derive exact Green’s function solutions for a geometrically thin, time-dependent accretion disk with magnetically driven winds, considering three types of inner boundary conditions—zero torque, zero mass accretion rate, and finite torque with finite mass accretion rate—applied both at the origin and at a finite inner radius.

In Section~\ref{sec:beq}, we describe the basic equations governing a thin accretion disk influenced by magnetically driven winds. Sections~\ref{sec:torzero}, \ref{sec:masszero}, and \ref{sec:SFrin} present the Green’s function solutions for three different inner boundary conditions: zero torque, zero mass accretion rate, and finite torque with finite accretion rate. In Section~\ref{sec:comp_par}, we compare how the mass accretion rate, wind mass-loss rate, and bolometric luminosity evolve over time under these conditions. In Section~\ref{sec:discus}, we apply our solution to the evolution of protoplanetary disks, examining how different boundary conditions affect their evolution and lifetime. We conclude with a summary of our main results in Section~\ref{sec:sumcon}.


%
\section{Basic Equations} 
\label{sec:beq}
%

We examine a time-dependent, one-dimensional accretion disk that is geometrically thin, optically thick, and axisymmetric, and launches a wind driven by magnetic stresses. Time-dependent models for such systems have previously been developed by Suzuki et al. (2016) \cite{2016A&A...596A..74S} in the context of protoplanetary disks, and by Tamilan et al. (2024, 2025) \cite{2024ApJ...975...94T,2025PTEP.2025h3E01T} for tidal disruption events (TDEs). A steady-state framework suited to X-ray binaries and active galactic nuclei (AGN) was later formulated by Tamilan et al. (2025) \cite{2024arXiv241100298T}. The present analysis follows the same foundational equations established in Tamilan et al. (2024, 2025) \cite{2024ApJ...975...94T, 2024arXiv241100298T}. For complete derivations, readers are referred to Section 2 and Appendix A of Tamilan et al. (2025). Here, we briefly summarize the governing conservation laws—mass and angular momentum—along with the key parameters defining the model.

The vertically integrated mass conservation equation is given by
\begin{equation}\label{eq:sigt}
	\frac{\partial \Sigma}{\partial t} 
	+ 
	\frac{1}{r}\frac{\partial}{\partial r}(r \Sigma v_r) + \dot{\Sigma}_{\rm w}  = 0,
\end{equation}
where $\dot{\Sigma}_{\rm w}$ is the vertical mass flux of the wind, $\Sigma = 2 H \rho$ is the surface density in the disk {with the disk's mass density $\rho$}, $H$ is the disk scale height and $v_r$ is the radial velocity. The angular momentum conservation equation is given by
\begin{equation}\label{eq:cylang}
	\frac{\partial}{\partial t}(r^2\Omega \Sigma) + \frac{1}{r} \frac{\partial}{\partial r} \left[r^2 \Sigma\left\{v_r r \Omega + \bar{\alpha}_{r\phi} c_s^2 \right\}\right] + r \left[\dot{\Sigma}_{\rm w}  r \Omega + \bar{\alpha}_{z\phi} \rho c_s^2\right] = 0,
\end{equation}
where $\bar{\alpha}_{r\phi}$ and $\bar{\alpha}_{z\phi}$ are introduced as parameters due to the MHD turbulence and disk winds (see Appendix A of \cite{2024arXiv241100298T}), and the Keplerian angular velocity is
\begin{equation}\label{eq:omega}
	\Omega = \sqrt{\frac{G M }{r^3}}
\end{equation}
with $G$ and $M$ being the Gravitational constant and the mass of the central object, respectively. Substituting equations (\ref{eq:sigt}) and (\ref{eq:omega}) in equation (\ref{eq:cylang}) results in 
\begin{equation}
	\label{eq:rsvr}
	r \Sigma v_r = -\frac{2}{r\Omega} 
	\left[\frac{\partial}{\partial r} 
	\left( 
	\bar{\alpha}_{r\phi} r^2 \Sigma c_s^2 
	\right) 
	+ 
	\frac{\bar{\alpha}_{z\phi}}{2} 
	\frac{r^2 \Sigma c_s^2}{H}
	\right].
\end{equation} 

As in Tamilan et al. (2025) \cite{2024arXiv241100298T}, we assume the magnetic braking parameter to be
\begin{equation}\label{eq:azphi}
	\bar{\alpha}_{z\phi} = \bar{\alpha}_{r\phi} \psi \frac{H}{r},
\end{equation}
where $\bar{\alpha}_{r\phi}$ is a constant, typically ranging from 0.01 to 0.1, and $\psi$ is a dimensionless free parameter. This formulation implies that $\bar{\alpha}_{z\phi}$ varies with radius and time. When $\psi$ is small, the vertical magnetic stress is weak, consistent with poloidal fields being relatively weak and magnetorotational instability (MRI) playing a central role in amplifying the field and driving outflows through magnetic pressure and reconnection \cite{2009ApJ...691L..49S}. As $\psi$ increases, stronger poloidal fields give rise to greater vertical stress, enabling more effective angular momentum extraction. In such cases, gas can either accrete onto the central object or be lifted along magnetic field lines and ejected in a wind via the magnetocentrifugal mechanism \cite{1982MNRAS.199..883B}.

Tabone et al. (2022) \cite{2022MNRAS.512.2290T} introduced the parameters $\alpha_{\rm SS}$ and $\alpha_{\rm DW}$ to describe radial and vertical turbulent stresses, respectively, in their disk-wind model. When compared with our notation, these relate as $\bar{\alpha}_{r\phi}=(3/2)\alpha_{\rm SS}$ and  $\bar{\alpha}_{z\phi}=(3/2)\alpha_{\rm DW}(H/r)$, leading to $\psi=\alpha_{\rm DW}/\alpha_{\rm SS}$, where equation~(\ref{eq:azphi}) was applied. Following Tabone et al. (2022) \cite{2022MNRAS.512.2290T}, the vertical mass flux is 
\begin{equation} \label{eq:dsigwdt}
	\dot{\Sigma}_{\rm w} = \frac{\psi \bar{\alpha}_{r\phi}}{2(\lambda-1)}\frac{c_s^2\Sigma}{r^2\Omega},
\end{equation}
where the magnetic lever arm parameter $\lambda$ represents the ratio of specific angular momentum carried away by the MHD disk wind at $r$ to the specific angular momentum of the disk at same radius. By substituting equations (\ref{eq:rsvr}) and (\ref{eq:dsigwdt}) in equation (\ref{eq:sigt}), the surface density evolution equation is given by
\begin{equation}\label{eq:sigevn}
	\frac{\partial \Sigma}{\partial t} - \frac{2}{\sqrt{GM}}\frac{1}{r}\frac{\partial}{\partial r}\left[\sqrt{r}\frac{\partial}{\partial r}\left\{\bar{\alpha}_{r\phi}r^2\Sigma c_s^2\right\}\right]-\frac{1}{\sqrt{GM}}\frac{\psi}{r}\frac{\partial}{\partial r}\left[\bar{\alpha}_{r\phi} r^{3/2}c_s^2\Sigma\right] + \frac{\psi \bar{\alpha}_{r\phi}}{2(\lambda-1)}\frac{c_s^2\Sigma}{r^2\Omega}=0.
\end{equation}

For a hydrodynamic disk, turbulent viscosity is given by $\nu = \alpha c_s H = \alpha c_s^2/\Omega$, where $\alpha = 2\bar{\alpha}_{r\phi} / 3 $ \cite{2024ApJ...975...94T}. This results in $\bar{\alpha}_{r\phi} c_s^2/\Omega = 3 \nu / 2$, and assuming the turbulent viscosity to be  \cite{1974MNRAS.168..603L,2011MNRAS.410.1007T}
\begin{equation}\label{eq:nur}
	\nu = \nu_{\rm c} r^{n},
\end{equation}
where $\nu_{\rm c}$ and $n$ are constants, we get 
\begin{equation}\label{eq:arphics2}
	\bar{\alpha}_{r\phi} c_s^2 = \frac{3\nu_{\rm c}}{2} \sqrt{GM}r^{-3/2+n}.
\end{equation}
We note that Tabone et al. (2022) \cite{2022MNRAS.512.2290T} adopted $\alpha_{\rm SS} c_s^2 \propto r^{-3/2+\gamma}$ and $\alpha_{\rm DW} c_s^2 \propto r^{-3/2+\gamma}$ in their model. This is consistent with our formulation for $\gamma = n$, $\bar{\alpha}_{r\phi} = (3/2)\alpha_{\rm SS}$, and $\psi = \alpha_{\rm DW}/\alpha_{\rm SS}$ (see above equation~\ref{eq:dsigwdt}).
By substituting equation (\ref{eq:arphics2}) in equation (\ref{eq:sigevn}), the surface density evolution equation is
\begin{equation}\label{eq:sigevn2}
	\frac{\partial \Sigma}{\partial t} - \frac{3 \nu_{\rm c}}{r}\frac{\partial}{\partial r}\left[\sqrt{r}\frac{\partial}{\partial r}\left\{r^{n+1/2}\Sigma\right\}\right]-\frac{3 \nu_{\rm c}}{2}\frac{\psi}{r}\frac{\partial}{\partial r}\left[r^{n}\Sigma\right] + \frac{3 \psi}{4(\lambda-1)} \nu_{\rm c}\frac{\Sigma}{r^{2-n}}=0.
\end{equation}
In the absence of wind, $\psi = 0$, equation (\ref{eq:sigevn2}) reduces to the surface density evolution equation given in Tanaka (2011) \cite{2011MNRAS.410.1007T}.
  
Assuming a separable ansatz of the form $\Sigma(r,t) = F(r) e^{-\Lambda t}$ \cite{2002apa..book.....F}, where $\Lambda$ is a real number and $F(r)$ is an arbitrary function of $r$, equation (\ref{eq:sigevn2}) reduces to
\begin{equation}\label{eq:sigevn3}
	r^2 \frac{\partial^2 F}{\partial r^2} + \left(2n+\frac{3}{2} + \frac{\psi}{2}\right) r \frac{\partial F}{\partial r} + \left[\frac{\Lambda}{3 \nu_{\rm c}} r^{2-n} + \left\{n\left(n+\frac{1}{2}+\frac{\psi}{2}\right) - \frac{\psi}{4(\lambda-1)}\right\}\right]F(r) = 0.
\end{equation}
With the choice of $\Lambda = 3 \nu_{\rm c}k^2$, the general solution of equation (\ref{eq:sigevn3}) is given by
\begin{equation}\label{eq:Frn}
	F(r) = r^{-(4n+1+\psi)/4} \left[C_1(k) J_{l}\left(\frac{k}{\beta} r^{\beta}\right)+ C_2(k) Y_{l}\left(\frac{k}{\beta} r^{\beta}\right)\right],
\end{equation} 
where $k$ is an arbitrary mode of the solution, $C_1(k)$ and $C_2(k)$ are the mode weights, $J_{l}$ and $Y_{l}$ are the Bessel functions of the first and second kinds, respectively, and
\begin{eqnarray}
	\beta &=& 1-\frac{n}{2}, \label{eq:delta}\\
	l &=& \frac{\sqrt{(1+\psi)^2+4\psi/(\lambda-1)}}{4\beta} \label{eq:l}
\end{eqnarray}
with $l>0$ indicating $\beta > 0$ and $n < 2$. Using equation (\ref{eq:Frn}) to obtain the surface density $\Sigma(r,t) = F(r) e^{-\Lambda t}$, and integrating the fundamental solution across all possible $k$-modes gives the solution of the surface density as
\begin{equation}\label{eq:sigsoln}
	\Sigma(r,t) = r^{-(4n+1+\psi)/4}\int_0^{\infty}\left[C_1(k) J_{l}\left(\frac{k}{\beta} r^{\beta}\right)+ C_2(k) Y_{l}\left(\frac{k}{\beta} r^{\beta}\right)\right] e^{-3 \nu_{\rm c}t k^2} \, \diff k.
\end{equation}
The mode-weighting functions $C_1(k)$ and $C_2(k)$ are determined by the boundary condition and the initial surface density profile $\Sigma(r, t =0)$. Our aim is to rewrite equation (\ref{eq:sigsoln}) in the Green’s function form given by \cite{2011MNRAS.410.1007T}
\begin{equation}\label{eq:greenfncn}
	\Sigma(r,t) = \int_{r_{\rm in}}^{\infty} \Sigma(r^{'},t=0) \mathcal{G}(r,r^{'},t) \, \diff r^{'},
\end{equation}
where $\mathcal{G}(r,r^{'},t)$ is the Green's function. Our strategy is to first use the boundary condition to establish a relationship between the mode weights $C_1(k)$ and $C_2(k)$. Next, we apply an appropriate integral transform to express the mode weights in terms of the initial surface density profile $\Sigma(r, t =0)$. By substituting the resulting expressions for the mode weights into equation (\ref{eq:sigsoln}) and integrating over all modes $k$, we obtain the Green's function.

\subsection{Mass accretion and wind rates, and bolometric luminosity}

With the derived time-dependent solution for the surface density, we can estimate the disk physical quantities that are mass accretion and wind rates, and bolometric luminosity.

Using equations (\ref{eq:rsvr}), (\ref{eq:azphi}) and (\ref{eq:arphics2}), the mass accretion rate, $\dot{M}=-2\pi r \Sigma v_r$, is given by
\begin{equation}\label{eq:mdotacc}
	\dot{M} = 6 \pi \nu_{\rm c} r^{(1-\psi)/2}\frac{\partial}{\partial r}\left[r^{n+(1+\psi)/2}\Sigma\right].
\end{equation} 
The wind mass loss rate rate is given by 
\begin{equation}\label{eq:mwind}
	\dot{M}_{\rm w} = \int_{r_{\rm in}}^{r} 2 \pi \dot{\Sigma}_{\rm w} r \, \diff r = \frac{3\pi \psi}{2(\lambda-1)} \nu_{\rm c} \int_{r_{\rm in}}^{r} r^{1-2\beta} \Sigma \, \diff r, 
\end{equation}
where we have used equations (\ref{eq:dsigwdt}) and (\ref{eq:arphics2}) for the derivation. The energy conservation equation for an accretion disk with magnetically driven wind is \cite{2024ApJ...975...94T,2024arXiv241100298T}
\begin{equation}\label{eq:ent}
	\dot{\Sigma}_{\rm w} \frac{r^2\Omega^2}{2} + Q_{\rm rad} = \frac{3}{2}\bar{\alpha}_{r\phi} \Omega \Sigma c_s^2 + \frac{1}{2} \bar{\alpha}_{z\phi} r \Omega^2 \Sigma c_s,
\end{equation}
where the first term on left hand side indicates the kinetic energy flux carried by wind, $Q_{\rm rad}$ is the radiative flux, and the right hand side represent the heating flux due to turbulent viscosity and magnetic braking. Using equations (\ref{eq:azphi}), (\ref{eq:dsigwdt}) and (\ref{eq:arphics2}) in equation (\ref{eq:ent}), the radiative flux is given by
\begin{equation}
	Q_{\rm rad} = \frac{3}{4}\left[3+\frac{\psi(2\lambda-3)}{2(\lambda-1)}\right] \nu_{\rm c} \frac{G M}{r^{3-n}}\Sigma.
\end{equation}
According to the Stefan-Bolzmann law, the radiative flux emitted from the disk surface is $Q_{\rm rad}=2\sigma T_{\rm eff}^4$, where $T_{\rm eff}$ is the surface temperature of the disk and $\sigma$ is the Stefan-Boltzmann constant. The bolometric luminosity is then given by 
\begin{eqnarray}\label{eq:bollum}
	L
	= \int_{r_{\rm in}}^{r_{\rm out}} Q_{\rm rad} 2\pi r\, \diff r = \frac{3\pi}{2}\left[3+\frac{\psi(2\lambda-3)}{2(\lambda-1)}\right] \nu_{\rm c} G M \int_{r_{\rm in}}^{r_{\rm out}} \frac{\Sigma}{r^{2-n}} \, \diff r,
\end{eqnarray}
where $r_{\rm out}$ indicates the outer radius of the disk. 

In the subsequent sections, we derive the Green's function for three different boundary conditions imposed at the inner radius of the disk, corresponding to the following cases:
\begin{enumerate}
	\item {\bf Case I:} Zero torque at the inner boundary of the disk.
	\item {\bf Case II:} Zero mass accretion rate at the inner boundary of the disk.
	\item {\bf Case III:} Finite torque and finite mass accretion rate at the inner boundary of the disk.
\end{enumerate}
The specific conditions required for the surface density in each case are discussed in detail in the respective sections.

%
\section{Case I: Zero torque inner boundary condition} 
\label{sec:torzero}
%

In this section, we derive the Green's function and the corresponding surface density solution for a disk with a zero-torque boundary condition at the inner edge, $r_{\rm in}$. The radial torque due to turbulent viscosity is given by $ 3\pi \nu \Sigma r^2 \Omega$ \cite{2002apa..book.....F}. From equation (\ref{eq:cylang}), the rate at which the wind extracts angular momentum from the disk per unit area is $r \left[\dot{\Sigma}_{\rm w}  r \Omega + \bar{\alpha}_{z\phi} \rho c_s^2\right] = \lambda r^2 \Omega \dot{\Sigma}_{\rm w}$, where we have used equation (\ref{eq:dsigwdt}). Thus, the wind torque per unit area is $\lambda r^2 \Omega \dot{\Sigma}_{\rm w}$. The total torque at any radius $r$ is the sum of the radial and vertical torques, which is approximated as $\mathcal{T} = 3\pi \nu \Sigma r^2 \Omega + r^2 \lambda r^2 \Omega \dot{\Sigma}_{\rm w} \propto \left(1+\psi/2\right) \sqrt{r} \nu \Sigma \propto \left(1+\psi/2\right) r^{n+1/2}\Sigma$.
Note that an additional factor of $r^2$ is included to account for the area element when calculating the net wind torque.  In case of disk's inner radius $r_{\rm in}>0$, the zero-torque boundary condition indicates $\Sigma(r_{\rm in}) = 0$. Using equation (\ref{eq:sigsoln}), the total torque is given
\begin{equation}\label{eq:torque1}
	\mathcal{T} \propto \int_{0}^{\infty} \left[C_1(k) r^{(1-\psi)/4} J_{l}\left(\frac{k}{\beta} r^{\beta}\right)+ C_2(k) r^{(1-\psi)/4}Y_{l}\left(\frac{k}{\beta} r^{\beta}\right)\right] e^{-3 \nu_{\rm c}t k^2} \, \diff k,
\end{equation}
indicating that the zero torque at $r_{\rm in}$ corresponds to 
\begin{equation}\label{eq:torbnd}
	C_1(k) r_{\rm in}^{(1-\psi)/4} J_{l}\left(\frac{k}{\beta} r_{\rm in}^{\beta}\right)+ C_2(k) r_{\rm in}^{(1-\psi)/4}Y_{l}\left(\frac{k}{\beta} r_{\rm in}^{\beta}\right) = 0.
\end{equation}

\subsection{Inner edge at the origin, $r_{\rm in} = 0$}
\label{sec:zeror0}

In the limit of $r \rightarrow 0$, $J_{l}\left((k/\beta) r^{\beta}\right) \propto r^{\beta l} $ and $Y_{l}\left((k/\beta) r^{\beta}\right) \propto r^{-\beta l} $, where it is to note from equation (\ref{eq:l}) that $\beta l = \sqrt{(1+\psi)^2+4\psi/(\lambda-1)} / 4  > 0$. Thus, as $r \rightarrow 0$, $J_{l}\left((k/\beta) r^{\beta}\right)$ approaches zero, whereas $Y_{l}\left((k/\beta) r^{\beta}\right)$ diverges. Consequently, for an inner boundary at $r_{\rm in} = 0$, the solution satisfies the zero-torque condition if $C_2(k) = 0$. Then, equation (\ref{eq:sigsoln}) reduces to
\begin{equation}\label{eq:sigsoln1}
	\Sigma(r,t) =  r^{-(4n+1+\psi)/4}\int_0^{\infty} C_1(k) J_{l}\left(\frac{k}{\beta} r^{\beta}\right) e^{-3 \nu_{\rm c}t k^2} \, \diff k.
\end{equation}


Applying the Hankel transformation (equations~\ref{eq:Hankel} and~\ref{eq:Hankel1} in Appendix~\ref{sec:inttrans}) to equation~(\ref{eq:sigsoln1}) at $t = 0$ gives
\begin{equation}\label{eq:ckkn}
	\frac{C_1(k)}{k} = \frac{1}{\beta}\int_{0}^{\infty} r^{'(5+\psi)/4} \Sigma(r^{'},t = 0) J_{l}\left(\frac{k}{\beta} r^{'\beta}\right) \, \diff r^{'}.
\end{equation}
Substituting equation (\ref{eq:ckkn}) in equation (\ref{eq:sigsoln1}) results in
\begin{equation}\label{eq:sigsoln3}
	\Sigma(r,t) = \int_0^{\infty} \Sigma(r^{'},t = 0) \beta^{-1}r^{-(4n+1+\psi)/4}r^{'(5+\psi)/4} \int_0^{\infty} J_{l}\left(\frac{k}{\beta} r^{'\beta}\right) J_{l}\left(\frac{k}{\beta} r^{\beta}\right) e^{-3 \nu_{\rm c}t k^2} k \, \diff k \, \diff r^{'}.
\end{equation}
Comparing equation (\ref{eq:sigsoln3}) with equation (\ref{eq:greenfncn}) gives the Green's function as
\begin{eqnarray}
	\mathcal{G}(r,r^{'},t) &&= \frac{1}{\beta} r^{-(4n+1+\psi)/4}r^{'(5+\psi)/4} \int_0^{\infty} J_{l}\left(\frac{k}{\beta} r^{'\beta}\right) J_{l}\left(\frac{k}{\beta} r^{\beta}\right) e^{-3 \nu_{\rm c}t k^2} k \, \diff k, \nonumber \\
	&&=\frac{1}{6 \beta \tau(r)}r^{-(9+\psi)/4}r^{'(5+\psi)/4} I_{l}\left[\frac{1}{6 \beta^2 \tau(r)}\left(\frac{r^{'}}{r}\right)^{\beta}\right] \exp\left(-\frac{1+(r^{'} / r)^{2\beta}}{12 \beta^2 \tau(r)}\right). \label{eq:gfunz}
\end{eqnarray} 
Here, $\tau(r) =t / t_{\nu}(r)$ and $t_{\nu}(r) = r^2/\nu$ is the local viscous timescale at radius $r$. In the absence of the wind, $\psi =0$, equation (\ref{eq:gfunz}) is identical to the Green's function solution of Tanaka (2011) \cite{2011MNRAS.410.1007T}.

While the Green’s function framework permits computation of $\Sigma(r,t)$ for any given initial surface density profile, it is particularly insightful to examine the scenario where the initial surface density is represented by a Dirac delta function \cite{2002apa..book.....F,2011MNRAS.410.1007T},
\begin{equation}\label{eq:inisd}
	\Sigma(r, t= 0) = \Sigma_0\delta(r-r_0) r_0,
\end{equation}
for which the solution is (by definition) the Green’s function itself. The equation (\ref{eq:inisd}) represents an infinitesimally narrow ring of total mass concentrated at $r_0$ and $\Sigma_0$ is a normalization constant. The mass of the initial disk is given by
\begin{equation}\label{eq:mdini}
M_{d,0} = \int \Sigma(r, t= 0) 2 \pi r \, dr = 2 \pi r_0^2 \Sigma_0.
\end{equation}

Using equations (\ref{eq:inisd}) and (\ref{eq:gfunz}) in equation (\ref{eq:greenfncn}) results in the surface density 
\begin{multline}\label{eq:sigzt0}
	\Sigma(r,t)=\frac{\Sigma_0}{6 \beta}\left(\frac{t}{t_{\nu}(r_0)}\right)^{-1}\left(\frac{r}{r_0}\right)^{-n-(1+\psi)/4} I_{l}\left[\frac{1}{6 \beta^2}\left(\frac{t}{t_{\nu}(r_0)}\right)^{-1}\left(\frac{r}{r_0}\right)^{\beta}\right]\\ \exp\left[-\frac{1}{12\beta^2}\left(\frac{t}{t_{\nu}(r_0)}\right)^{-1}\left\{1+\left(\frac{r}{ r_0}\right)^{2\beta}\right\}\right],
\end{multline}
where $t_{\nu}(r_0)$ denotes the viscous timescale evaluated at the reference radius $r_0$. In Appendix~\ref{sec:appcomp}, we compare the asymptotic solution of our surface density profile with that of Tabone et al. (2022)\cite{2022MNRAS.512.2290T}. We use the surface density given by equation (\ref{eq:sigzt0}) in equation (\ref{eq:mdotacc}) to compute the mass accretion rate $\dot{M}$, and in equation (\ref{eq:mwind}) to calculate the wind mass loss rate $\dot{M}_{\rm w}$. 

%
\begin{figure}
	\centering
	\subfigure[]{\includegraphics[scale = 0.63]{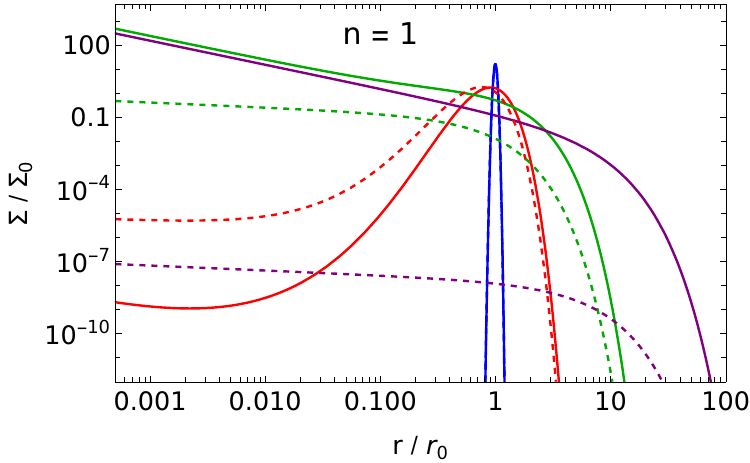}}
	\subfigure[]{\includegraphics[scale = 0.5]{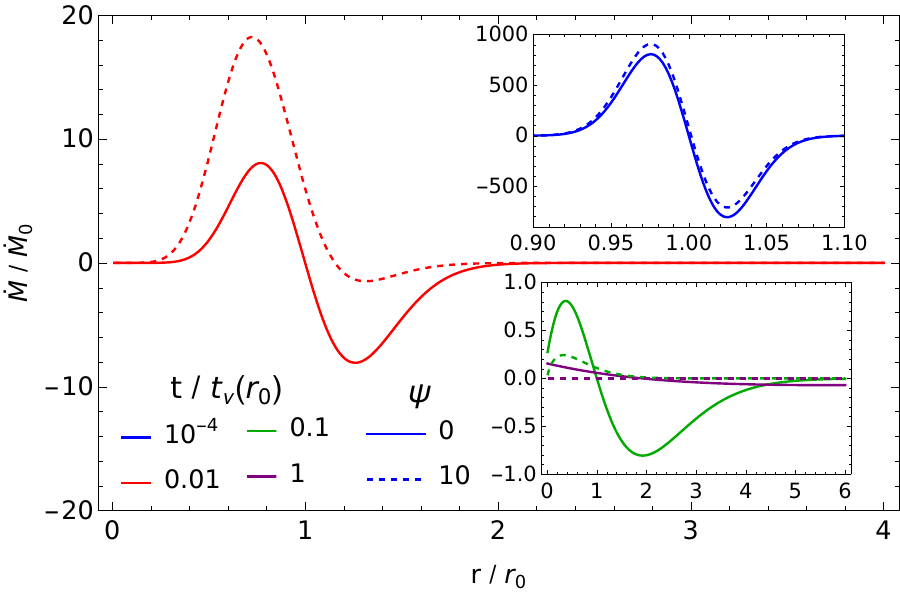}}
	\subfigure[]{\includegraphics[scale = 0.63]{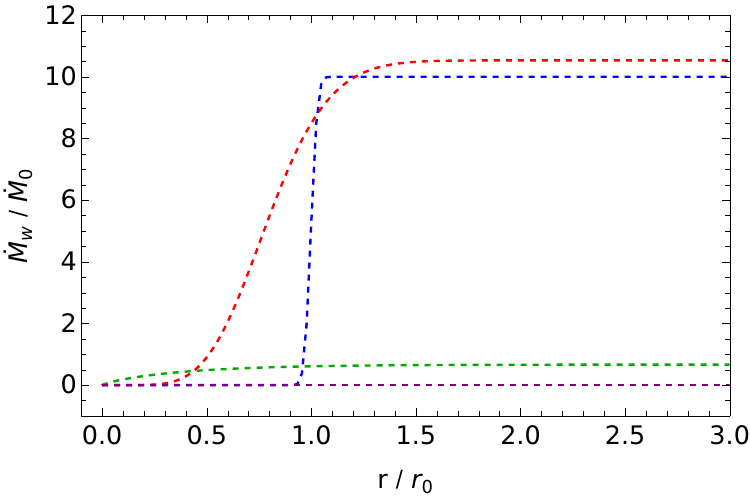}}
	\caption{
		Radial profiles of the surface density $\Sigma$ (panel a), mass accretion rate (panel b), and integrated wind mass-loss rate (panel c) for a disk with a zero-torque boundary condition at the inner edge ($r_{\rm in} = 0$). Results are shown for two values of the wind-driving parameter: $\psi = 0$ (solid lines; no wind) and $\psi = 10$ (dashed lines; with wind). The blue, red, green, and purple lines correspond to $t / t_{\nu}(r_0) = 10^{-4}$, $0.01$, $0.1$, and $1$, respectively. The magnetic lever arm parameter is fixed at $\lambda = 3/2$. The quantities $\Sigma_0$, $r_0$, and $\dot{M}_0 = 3 \pi \nu(r_0)\Sigma_0$ are arbitrary normalization constants. In panel (a), the surface density profiles for $\psi = 0$ (blue solid) and $\psi = 10$ (blue dashed) coincide at $t = 10^{-4} t_{\nu}(r_0)$. Note that in panel (c), solid lines are absent because $\psi = 0$ corresponds to the no-wind case.}
	\label{fig:ztrin0} 
\end{figure}
%


Figure~\ref{fig:ztrin0} illustrates the radial evolution of the surface density $\Sigma$, the mass accretion rate $\dot{M}$, and the outwardly integrated wind mass loss rate $\dot{M}_{\rm w}$ for disks without wind ($\psi = 0$, solid lines) and with wind ($\psi = 10$, dashed lines). The surface density is normalized by $\Sigma_0$, while the accretion and wind rates are normalized by 
\begin{eqnarray}\label{eq:mdot0}
	\dot{M}_0 &=& 3 \pi \nu(r_0) \Sigma_0 = \frac{3}{2} \frac{M_{d,0}}{t_{\nu}(r_0)}\nonumber\\
	&\simeq& 0.015~{\rm M_{\odot}yr^{-1}} \left(\frac{M_{d,0}}{M_{\odot}}\right)  \left(\frac{t_{\nu}(r_0)}{100~{\rm yr}}\right)^{-1}.
\end{eqnarray}
Panel (a) shows the radial evolution of the surface density over time, demonstrating viscous spreading of the disk gas. The initial surface density is described by a Dirac delta function centered at $r = r_0$, which is discontinuous and diverges at $r_0$, representing all the disk mass concentrated in an infinitesimally narrow ring. As such, it cannot be shown on a continuous radial scale, and the profiles are plotted from $t = 10^{-4} t_{\nu}(r_0)$ onward. However, the surface density profiles for $\psi = 0$ and $\psi = 10$ are identical at $t = 10^{-4} t_{\nu}(r_0)$, as the disk has not yet spread significantly from its initial distribution. The disk size increases due to the outward transport of angular momentum, while the expansion is reduced when a wind is present (for $\psi = 10$), as the wind removes part of the disk’s angular momentum. Panel (b) shows the mass accretion rate, which is higher in the presence of a wind at early times because the vertical magnetic stress associated with the wind removes angular momentum from the disk. This, combined with radial turbulent stress, enhances angular momentum transport and accelerates the inward gas flow. The combined effect of accretion and wind leads to a more rapid decline in surface density compared to the windless case. Consequently, at late times, both the surface density and accretion rate are lower in the wind-driven disk. As the surface density decreases, the vertical mass flux (equation~\ref{eq:dsigwdt}) also diminishes, resulting in a reduced mass-loss rate. Panel (c) shows the radial profile of the wind mass-loss rate, which increases at early times but drops significantly at later times.

We analyze the late-time behavior of the disk, where $t > t_{\nu}(r_0)$, in the inner region of the disk ($r < r_0$), such that $(t/t_{\nu}(r_0))^{-1}(r/r_0)^{\beta} \ll 1$. Under this condition, the asymptotic behavior of equation (\ref{eq:sigzt0}) is
\begin{multline}\label{eq:sigztl}
	\Sigma \left(r<r_0, t>t_{\nu}(r_0)\right) \sim \frac{2\beta \Sigma_0}{(12 \beta^2)^{1+l} \Gamma(1+l)} \left(\frac{t}{t_{\nu}(r_0)}\right)^{-(1+l)} \left(\frac{r}{r_0}\right)^{-n+\beta l-\left(1+\psi\right)/4},
\end{multline}
where we have used $I_{l}(z) \sim [\Gamma(1+l)]^{-1} (z/2)^{l} $ for the argument $z \ll 1$. Since the viscous timescale is $t_{\nu}(r) \propto r^{2 - n} \propto r^{2\beta}$, the viscous timescale at $r < r_0$ is shorter than that at $r_0$, indicating that at late times $t$, we have $t > t_{\nu}(r)$ for $r < r_0$. By substituting equation (\ref{eq:sigztl}) in equation (\ref{eq:mdotacc}), the mass accretion rate in the inner region of the disk at late times is given by
\begin{multline}\label{eq:mdotztor}
\dot{M}\left(r<r_0, t>t_{\nu}(r_0)\right) \sim \frac{4 \dot{M}_0 }{(12\beta^2)^{1+l}\Gamma(1+l)} \left[\beta l +\frac{1+\psi}{4}\right] \left(\frac{t}{t_{\nu}(r_0)}\right)^{-(1+l)}  \left(\frac{r}{r_0}\right)^{\beta l- \left(1+\psi\right)/4}.
\end{multline}
The mass accretion rate is $\dot{M} \propto t^{-1-l} r^{\beta l-(1+\psi)/4}$. When the wind is absent, $\psi=0$, the parameter $l$ reduces to $1/(4-2n)$. For example, choosing $n=1$, which corresponds to a sound speed profile $c_s^2 \propto r^{-1/2}$ (see equation \ref{eq:arphics2}), the mass accretion rate declines as $\dot{M} \propto t^{-3/2}$, consistent with the viscous disk evolution described by Lynden-Bell \& Pringle (1974) \cite{1974MNRAS.168..603L}. When the wind is present ($\psi > 0$), $l$ increases resulting in a steeper decline in the mass accretion rate. 

By substituting equation (\ref{eq:sigztl}) in equation (\ref{eq:mwind}), the wind mass loss rate from the inner region of the disk at the late times is given by
\begin{multline}\label{eq:mwztor}
\dot{M}_{\rm w}\left(r<r_0, t>t_{\nu}(r_0)\right)\sim \frac{4\dot{M}_0}{12^{1+l} \beta^{1+2l} \Gamma(1+l)} \left[\beta l +\frac{1+\psi}{4}\right]\left(\frac{t}{t_{\nu}(r_0)}\right)^{-(1+l)}  \left(\frac{r}{r_0}\right)^{\beta l- \left(1+\psi\right)/4}
\end{multline}
for $\psi > 0$. The wind mass loss rate follows the same radial and time profile as the mass accretion rate. Using equation (\ref{eq:sigztl}) in equation (\ref{eq:bollum}), the bolometric luminosity is given by
\begin{multline}\label{eq:bolzt0}
	L \sim 2 \left[3+\frac{\psi(2\lambda-3)}{2(\lambda-1)}\right] \frac{\beta L_0}{(12\beta^2)^{1+l}\Gamma(1+l)} \left(\frac{t}{t_{\nu}(r_0)}\right)^{-(1+l)} \\  \int_{0}^{r/r_0} \left(\frac{r^{'}}{r_0}\right)^{\left(\sqrt{(1+\psi)^2+4\psi/(\lambda-1)} - 9-\psi\right)/4} \, \diff\left(\frac{r^{'}}{r_0}\right),
\end{multline}
where $L_0$ is the luminosity normalization constant taken to be
\begin{eqnarray}\label{eq:lum0}
	L_0 &=& \frac{G M \dot{M}_0}{2 r_0} = \frac{3}{4} \frac{G M M_{d,0}}{r_0 t_{\nu}(r_0)} \nonumber \\
	&\simeq& 4.28 \times 10^{43}~{\rm erg~s^{-1}}~\left(\frac{M_{d,0}}{M_{\odot}}\right) \left(\frac{r_0}{10 r_g}\right) ^{-1} \left(\frac{t_{\nu}(r_0)}{100~{\rm yr}}\right)^{-1},
\end{eqnarray}
where $r_g = G M / c^2$ is the gravitational radius. Thus, the time evolution of the bolometric luminosity in the inner region of the disk follows the mass accretion, and diverges near the disk center ($r_{\rm in} = 0$) for $\psi= 0$ and when $\left(\sqrt{(1+\psi)^2 + 4\psi/(\lambda - 1)} - 9 - \psi\right)/4 \leq -1$, which corresponds to $\lambda \geq (6 + 3\psi)/(6 + 2\psi)$ for $\psi > 0$. Equations (\ref{eq:sigztl}) to (\ref{eq:bolzt0}) are valid near the inner edge of the disk at late times ($t > t_{\nu}(r_0)$). 


%
\begin{figure}
	\centering
	\subfigure[]{\includegraphics[scale = 0.63]{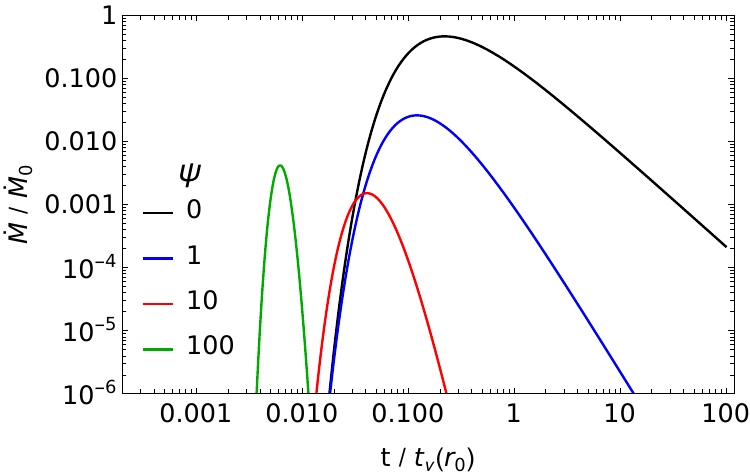}}
	\subfigure[]{\includegraphics[scale = 0.61]{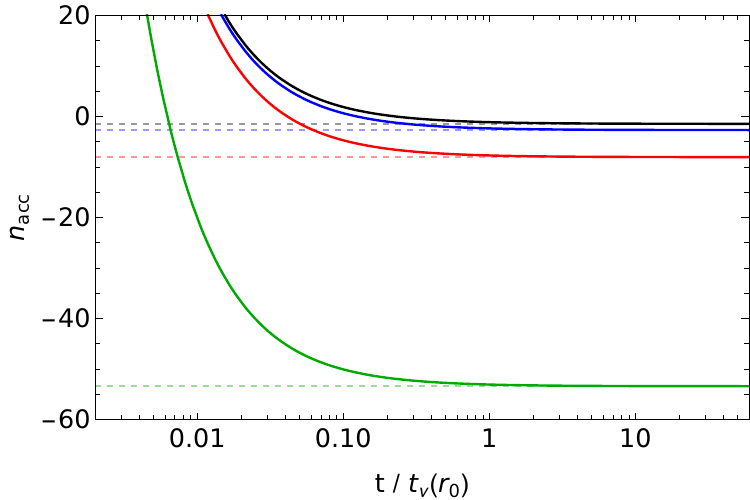}}
	\subfigure[]{\includegraphics[scale = 0.63]{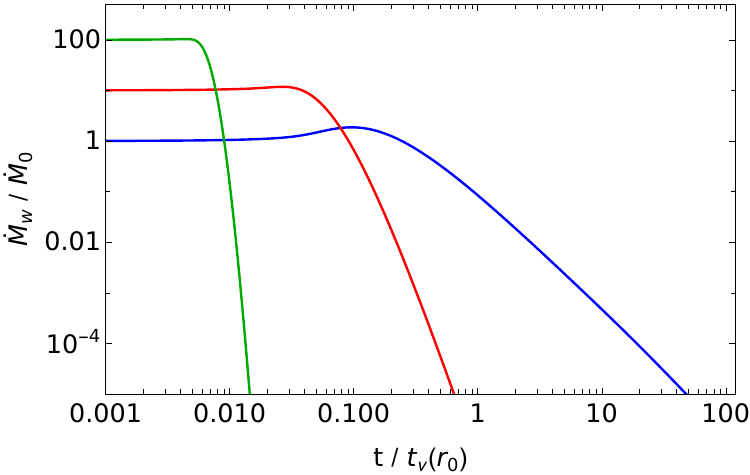}}
	\subfigure[]{\includegraphics[scale = 0.61]{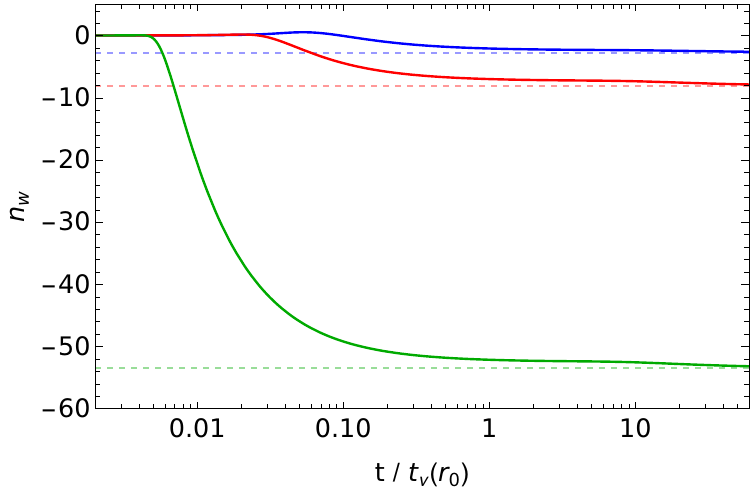}}
	\caption{
		The time evolution of the mass accretion rate, measured at $r = 10^{-5} r_0$ (inner region), and its corresponding slope, $n_{\rm acc} = \diff \ln \dot{M} / \diff \ln t$, are shown in panels (a) and (b), respectively, for a disk with a zero-torque boundary condition at the inner edge ($r_{\rm in} = 0$). Panels (c) and (d) present the total wind mass loss rate, integrated up to $r = 100 r_0$, and its slope, $n_{\rm w} = \diff \ln \dot{M}_{\rm w} / \diff \ln t$, respectively.  Different colors represent different values of $\psi$, with $\psi = 0$ corresponding to a disk without wind. The dashed lines in panels (b) and (d) indicate a slope of $-(1 + l)$, as predicted by equation~(\ref{eq:mdotztor}) for the late-time behavior of the mass accretion rate.
	}
	\label{fig:ztrin0_1} 
\end{figure}
%

In Figure~\ref{fig:ztrin0_1}, we show the time evolution of the mass accretion rate and the wind mass loss rate, computed by substituting equation~(\ref{eq:sigzt0}) into equations~(\ref{eq:mdotacc}) and (\ref{eq:mwind}), respectively. Panel (a) shows the mass accretion rate at $r = 10^{-5} r_0$, representing the inner disk region. The presence of wind leads to an earlier peak in the accretion rate due to enhanced radial inflow driven by vertical magnetic stress. As $\psi$ increases, the vertical stress strengthens, further accelerating the inflow and shortening the time to peak accretion. Panel (c) displays the total wind mass loss rate integrated up to $r = 100 r_0$, which increases with $\psi$ at early times, indicating more rapid mass loss. This results in a faster decline of both accretion and wind rates at late times. Panels (b) and (d) show the corresponding time evolution of the slopes, $n_{\rm acc} = \diff \ln \dot{M} / \diff \ln t$ and $n_{\rm w} = \diff \ln \dot{M}_{\rm w} / \diff \ln t$, which both approach the asymptotic value of $-(1 + l)$ at late times, in agreement with the predictions of equations~(\ref{eq:mdotztor}) and (\ref{eq:mwztor}).

In realistic astrophysical scenarios, accretion disks do not extend all the way to the center; instead, they are typically truncated at a physically meaningful inner boundary, such as the radius of the innermost stable circular orbit in the case of disks surrounding black holes.

\subsection{Inner edge at the finite radius, $r_{\rm in} > 0$}
\label{sec:ztorrinf}

In this case of finite inner radius, zero torque boundary condition results in $\Sigma = 0$ at $r_{\rm in}$. Equation (\ref{eq:torbnd}) gives the relation between $C_1(k)$ and $C_2(k)$ as
\begin{equation}
	C_2(k) = - C_{1}(k) \frac{J_{l}\left(\frac{k}{\beta} r_{\rm in}^{\beta}\right)}{Y_{l}\left(\frac{k}{\beta} r_{\rm in}^{\beta}\right)}.
\end{equation}
Substituting this relation in equation (\ref{eq:sigsoln}) gives 
\begin{equation}\label{eq:sigsolfn1}
	\Sigma(r,t) = r^{-(4n+1+\psi)/4}\int_0^{\infty}\frac{C_1(k)}{Y_{l}\left(\frac{k}{\beta} r_{\rm in}^{\beta}\right)}\left[J_{l}\left(\frac{k}{\beta} r^{\beta}\right)Y_{l}\left(\frac{k}{\beta} r_{\rm in}^{\beta}\right) - J_{l}\left(\frac{k}{\beta} r_{\rm in}^{\beta}\right)Y_{l}\left(\frac{k}{\beta} r^{\beta}\right)\right] e^{-3 \nu_{\rm c} t k^2} \, \diff k.
\end{equation}
By taking $k_1= (k /\beta) r_{\rm in}^{\beta}$, $r = r_{\rm in}x^{1/\beta}$ and $t=0$, and applying Weber integral transform pair (equations \ref{eq:webint} and \ref{eq:webint1} in Appendix \ref{sec:inttrans}) yields
\begin{multline}\label{eq:c1kft}
	\frac{C_1(k_1 \beta r_{\rm in}^{-\beta})(J_{l}^2(k_1)+Y_{l}^2(k_1))}{k_1 Y_{l}(k_1)} \\= \int_1^{\infty} \frac{r_{\rm in}^{\beta}}{\beta} (r_{\rm in} x^{'1/\beta})^{(4n+1+\psi)/4}\Sigma(x^{'},t = 0)  \left[J_{l}\left(k_1 x^{'}\right)Y_{l}(k_1) - J_{l}(k_1)Y_{l}\left(k_1 x^{'}\right)\right] x^{'} \, \diff x^{'}.
\end{multline}
Using $x = (r/r_{\rm in})^{\beta}$ (see above equation \ref{eq:c1kft}), and substituting equation (\ref{eq:c1kft}) in equation (\ref{eq:sigsolfn1}) gives
\begin{multline}
	\Sigma(r,t)=\int_{r_{\rm in}}^{\infty}\Sigma(r^{'},t=0) \frac{\beta}{r_{\rm in}^{2\beta}} r^{-(4n+1+\psi)/4}r^{'(5+\psi)/4} \int_0^{\infty} \left[J_{l}\left(k_1 (r^{'}/r_{\rm in})^{\beta}\right)Y_{l}(k_1) \right. \\ \left.- J_{l}(k_1)Y_{l}\left(k_1 (r^{'}/r_{\rm in})^{\beta}\right)\right] \left[\frac{J_{l}\left(k_1 (r/r_{\rm in})^{\beta}\right)Y_{l}(k_1) - J_{l}(k_1)Y_{l}\left(k_1 (r/r_{\rm in})^{\beta}\right)}{J_{l}^2(k_1)+Y_{l}^2(k_1)}\right]\\ e^{-3 \nu_{\rm c} \beta^2 r_{\rm in}^{-2\beta} t k_1^2} k_1 \, \diff k_1 \, \diff r^{'}.
\end{multline}
Following equation (\ref{eq:greenfncn}), the Green's function is 
\begin{multline}\label{eq:greensolfn}
	\mathcal{G}(r,r^{'},t) = \frac{\beta}{r_{\rm in}} \left(\frac{r}{r_{\rm in}}\right)^{-(4n+1+\psi)/4}\left(\frac{r^{'}}{r_{\rm in}}\right)^{(5+\psi)/4} \int_0^{\infty} \left[J_{l}\left(k_1 (r^{'}/r_{\rm in})^{\beta}\right)Y_{l}(k_1) \right. \\ \left. - J_{l}(k_1)Y_{l}\left(k_1 (r^{'}/r_{\rm in})^{\beta}\right)\right] \left[\frac{J_{l}\left(k_1 (r/r_{\rm in})^{\beta}\right)Y_{l}(k_1) - J_{l}(k_1)Y_{l}\left(k_1 (r/r_{\rm in})^{\beta}\right)}{J_{l}^2(k_1)+Y_{l}^2(k_1)}\right]e^{-3 \nu_{\rm c} \beta^2 r_{\rm in}^{-2\beta} t k_1^2} k_1 \, \diff k_1.
\end{multline} 
Equation (\ref{eq:greensolfn}) gives an exact expression for the Green’s function. There is no analytical solution for this Green's function, however, one can numerically estimate the time evolution of the disk for any given initial distribution of the surface density using this Green's function.


For $l = 1/2$, the Bessel function become easier to handle analytically with $J_{1/2}(x) = (2/\pi x)^{1/2} \sin x$ and $Y_{1/2}(x) = -(2/\pi x)^{1/2} \cos x$. In this case, the Green's function given by equation (\ref{eq:greensolfn}) reduces to 
\begin{multline}
\mathcal{G}(r,r^{'},t) =\frac{1}{2\sqrt{3\pi}r_{0}} \sqrt{\frac{t_{\nu}(r_{0})}{t}} \left(\frac{r}{r_0}\right)^{-(3n+3+\psi)/4}\left(\frac{r^{'}}{r_0}\right)^{(n+3+\psi)/4} \\ \left[\exp\left\{-\left(\left(\frac{r}{r_{0}}\right)^{\beta}-\left(\frac{r^{'}}{r_{0}}\right)^{\beta}\right)^2\frac{t_{\nu}(r_{0})}{12\beta^2 t}\right\} -\exp\left\{-\left(2\left(\frac{r_{\rm in}}{r_0}\right)^{\beta}-\left(\frac{r}{r_{0}}\right)^{\beta}-\left(\frac{r^{'}}{r_{0}}\right)^{\beta}\right)^2\frac{t_{\nu}(r_0)}{12\beta^2 t}\right\}\right].
\end{multline}


For any general $l$, the surface density derived by substituting the initial surface density profile given by equation (\ref{eq:inisd}) in (\ref{eq:greenfncn}) is given by
\begin{multline}\label{eq:sigztf}
	\Sigma(r,t) = \Sigma_0 \beta \left(\frac{r_0}{r_{\rm in}}\right)^{2\beta} \left(\frac{r}{r_{0}}\right)^{-(4n+1+\psi)/4} \int_0^{\infty} \left[J_{l}\left(k_1 (r_0/r_{\rm in})^{\beta}\right)Y_{l}(k_1) - J_{l}(k_1)Y_{l}\left(k_1 (r_0/r_{\rm in})^{\beta}\right)\right] \\ \left[\frac{J_{l}\left(k_1 (r_0/r_{\rm in})^{\beta}(r/r_{0})^{\beta} \right)Y_{l}(k_1) - J_{l}(k_1)Y_{l}\left(k_1 (r_0/r_{\rm in})^{\beta}(r/r_{0})^{\beta}\right)}{J_{l}^2(k_1)+Y_{l}^2(k_1)}\right]e^{-3 \beta^2 (r_0/r_{\rm in})^{2\beta} (t/t_{\nu}(r_0)) k_1^2} k_1 \, \diff k_1.
\end{multline}
We numerically estimate the surface density of the disk at a given radius and time using equation~(\ref{eq:sigztf}). By using equation~(\ref{eq:sigztf}) into equations~(\ref{eq:mdotacc}) and~(\ref{eq:mwind}), we numerically compute the mass accretion rate $\dot{M}$ and the wind mass loss rate $\dot{M}_{\rm w}$, respectively.

%
\begin{figure}
	\centering
	\subfigure[]{\includegraphics[scale = 0.53]{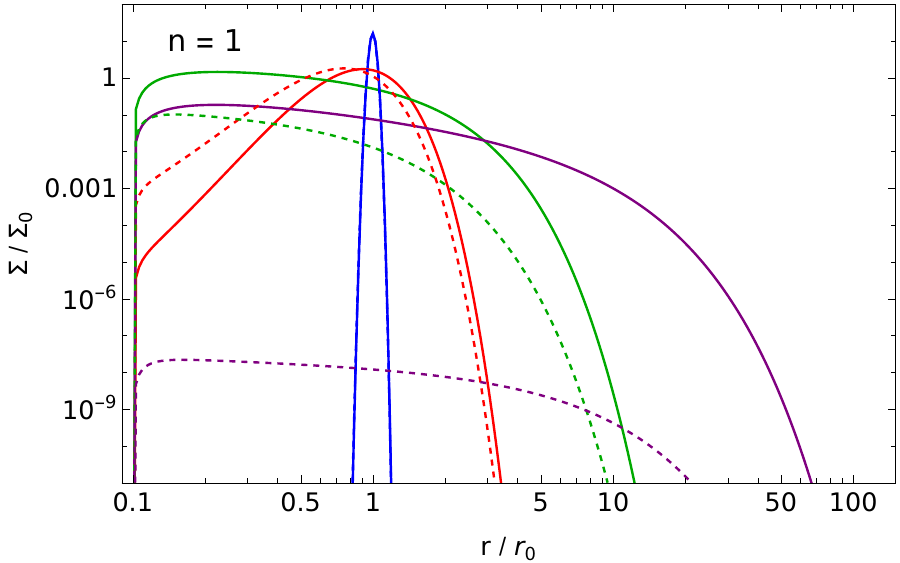}}
	\subfigure[]{\includegraphics[scale = 0.52]{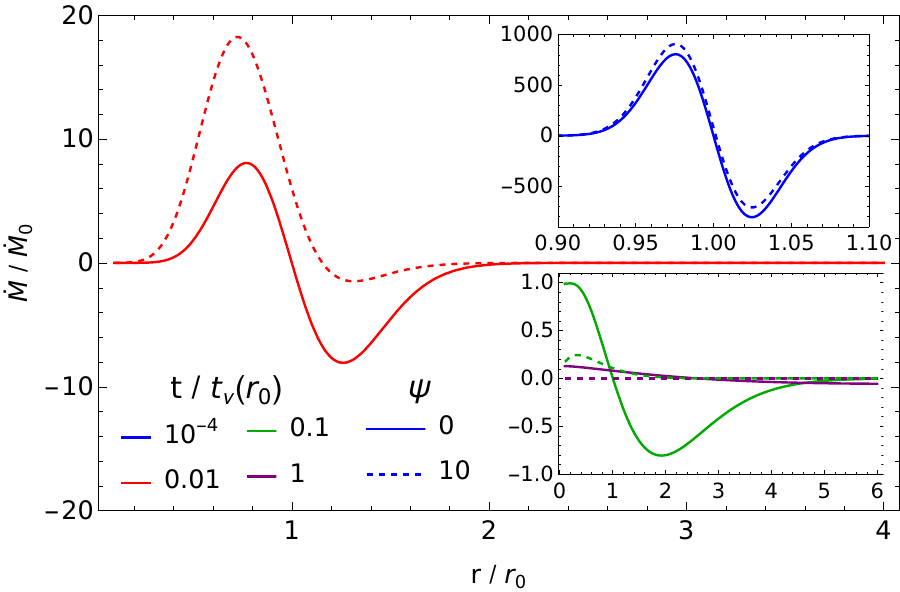}}
	\subfigure[]{\includegraphics[scale = 0.61]{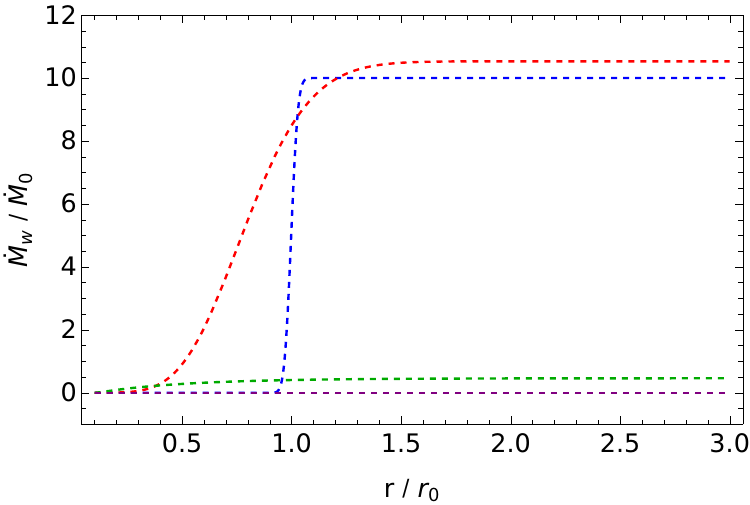}}
	\caption{
		The same format as in Figure~\ref{fig:ztrin0}, but for a disk with a zero-torque boundary condition at a finite inner edge. The inner radius of the disk is set to $r_{\rm in} = 0.1r_0$.
	}
	\label{fig:ztrinf} 
\end{figure}
%


Figure~\ref{fig:ztrinf} illustrates the radial evolution of the surface density $\Sigma$, the mass accretion rate $\dot{M}$, and the outwardly integrated wind mass loss rate $\dot{M}_{\rm w}$ for disks without wind ($\psi = 0$, solid lines) and with wind ($\psi = 10$, dashed lines). Panel (a) shows that the surface density vanishes at the inner edge, $r_{\rm in} = 0.1r_0$, consistent with the imposed boundary condition. Similar to the case shown in Figure~\ref{fig:ztrin0}, panel (b) shows that the mass accretion rate is higher at early times in the presence of wind. The wind enhances angular momentum transport, leading to an increased radial inflow and a higher accretion rate. The mass loss through both accretion and wind drives a rapid depletion of the disk mass. As a result, both the surface density and the mass accretion rate become lower at later times. This also leads to a substantial decline in the wind mass loss rate at late times, as shown in panel (c).


To analyze the late-time behavior of the disk, where $t > t_{\nu}(r_0) > t_{\nu}(r_{\rm in})$ and viscous timescale $t_{\nu}(r) \propto r^{2\beta}$, we observe that $(r_0/r_{\rm in})^{2\beta} (t/t_{\nu}(r_0)) = t/t_{\nu}(r_{\rm in}) > 1$. In this regime, the exponential term in the Green’s function suppresses the contribution from higher-order modes, and only modes with $k_1^2 \ll 1$ significantly influence the disk evolution. Assuming that $k_1 (r_0/r_{\rm in})^{\beta}$ is small, we can approximate the Bessel functions as $J_{l}(z) \sim [\Gamma(1+l)]^{-1}(z/2)^{l}$ and $Y_{l}(z) \sim -\left[\Gamma(l) / \pi\right](2/z)^{l}$ for an arbitrary small arguments $z < 1$. Under these assumptions, and focusing on the inner region of the disk where $r < r_0$, we derive an analytic expression for the surface density as
\begin{eqnarray}\label{eq:signzt}
	\Sigma(r< r_0, t> t_{\nu}(r_0)) &\simeq & \frac{3^{-1-l}\beta^{-1-2l}\Sigma_0}{2\Gamma(1+l)}\left(\frac{r_0}{r_{\rm in}}\right)^{-2\beta l} \left[\left(\frac{r_0}{r_{\rm in}}\right)^{2\beta l}-1\right] \left(\frac{t}{t_{\nu}(r_0)}\right)^{-(1+l)} \nonumber \\
	 && \left(\frac{r}{r_0}\right)^{\beta l - n -(1+\psi)/4} \left[1-\left(\frac{r_{\rm in}}{r}\right)^{2\beta l}\right].
\end{eqnarray}
The surface density vanishes at the inner edge of the disk, consistent with the imposed boundary condition. In the absence of the wind, $\psi=0$, $\beta l = 1/4$, which results in $\Sigma \propto r^{-n}\left[1-\left(r_{\rm in}/r\right)^{1/2}\right]$. This has been used extensively for solutions of accretion discs near zero-torque boundary surfaces \cite{1974MNRAS.168..603L,2002apa..book.....F}. Substituting equation (\ref{eq:signzt}) in equations (\ref{eq:mdotacc}) and (\ref{eq:mwind}) gives the mass accretion and loss rates near the inner edge as
\begin{eqnarray}\label{eq:mdotztf}
	\dot{M}(r< r_0, t> t_{\nu}(r_0)) =&& \frac{3^{-1-l}\beta^{-1-2l} \dot{M}_0}{\Gamma(1+l)} \left[1-\left(\frac{r_{\rm in}}{r_0}\right)^{2\beta l}\right] \left(\frac{t}{t_{\nu}(r_0)}\right)^{-(1+l)} \left(\frac{r}{r_0}\right)^{\beta l - (1+\psi)/4} \nonumber\\
	&& \left[\beta l +\frac{1+\psi}{4}+\left(\beta l - \frac{1+\psi}{4}\right)\left(\frac{r_{\rm in}}{r}\right)^{2\beta l}\right], \\
	\dot{M}_{\rm w}(r< r_0, t> t_{\nu}(r_0))=&& \frac{\left[\beta l + (1+\psi)/4\right]\dot{M}_0}{3^{1+l} \beta^{1+2l}\Gamma(1+l)} \left[1-\left(\frac{r_{\rm in}}{r_0}\right)^{2\beta l}\right] \left(\frac{t}{t_{\nu}(r_0)}\right)^{-(1+l)} \nonumber \\
	&& \left(\frac{r}{r_0}\right)^{\beta l - (1+\psi)/4} \left[1+\frac{4\beta l - (1+\psi)}{4 \beta l +1 +\psi} \left(\frac{r_{\rm in}}{r}\right)^{2\beta l} \right. \nonumber \\
	&&\left. -\frac{8\beta l}{4\beta l +1 + \psi}\left(\frac{r_{\rm in}}{r}\right)^{\beta l-(1+\psi)/4}\right],
\end{eqnarray}
respectively. It is noted that the mass accretion and mass loss rates at late times follow the same power-law time dependence as in the case with a zero-torque boundary at $r_{\rm in} = 0$, although their radial profiles differ near the inner edge. The bolometric luminosity, obtained using equations (\ref{eq:signzt}) and (\ref{eq:bollum}) and expressed in units of the mass accretion rate, is given by
\begin{multline}\label{eq:lumztf}
	L(r< r_0, t> t_{\nu}(r_0)) = \frac{G M \dot{M}(r_{\rm in})}{2 r_{\rm in}} \left[1-\frac{4\beta l +\psi+5}{8 \beta l} \left(\frac{r_{\rm in}}{r_t}\right)^{-\beta l+(5+\psi)/4}  \right. \\ \left. + \frac{5+\psi-4\beta l}{8 \beta l} \left(\frac{r_{\rm in}}{r_t}\right)^{\beta l+(5+\psi)/4}\right],
\end{multline}
where $r_t$ denotes the truncation radius below $r_0$, within which the disk has had sufficient time to reach the asymptotic regime. This expression shows that the bolometric luminosity follows the mass accretion rate near the inner edge of the disk. It is important to note that equations (\ref{eq:mdotztf}) to (\ref{eq:lumztf}) represent the asymptotic solution valid in the limit $t > t_{\nu}(r_0) > t_{\nu}(r_{\rm in})$, specifically for the region near the inner boundary. However, to accurately determine the solution at earlier times or across all radii, the surface density must be computed numerically using equation (\ref{eq:sigztf}).


%
\begin{figure}
	\centering
	\subfigure[]{\includegraphics[scale = 0.63]{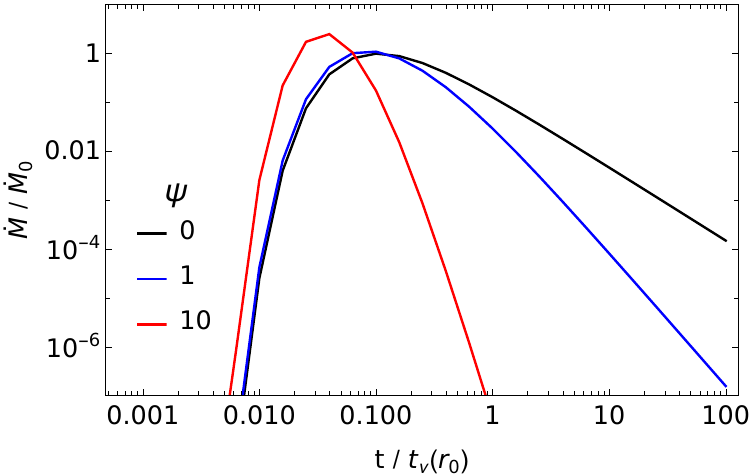}}
	\subfigure[]{\includegraphics[scale = 0.61]{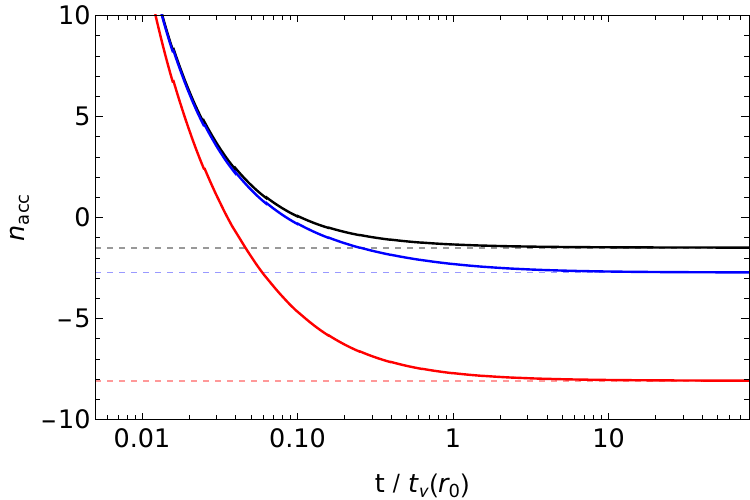}}
	\subfigure[]{\includegraphics[scale = 0.63]{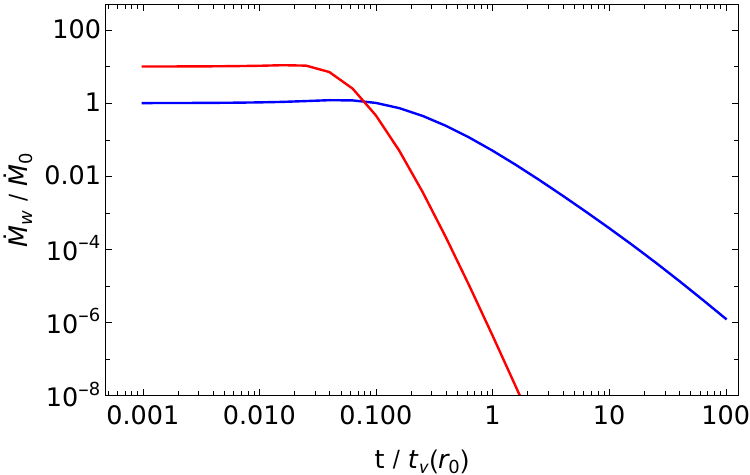}}
	\subfigure[]{\includegraphics[scale = 0.61]{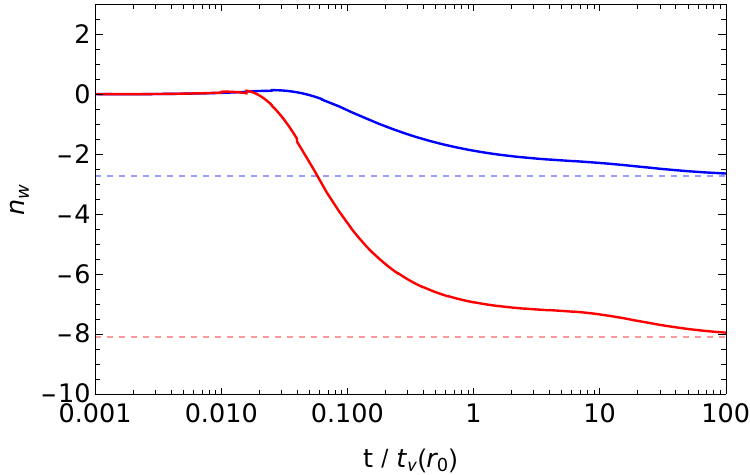}}
	\subfigure[]{\includegraphics[scale = 0.63]{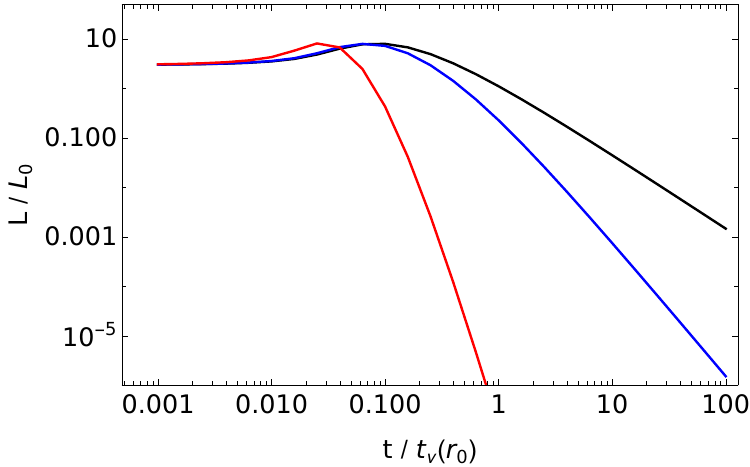}}
	\subfigure[]{\includegraphics[scale = 0.61]{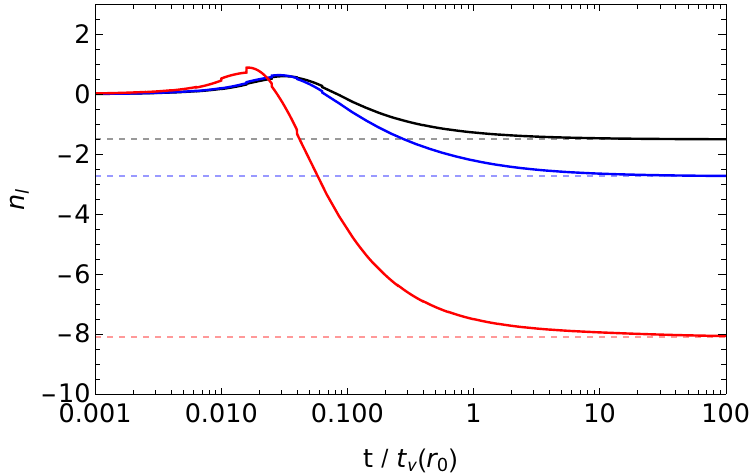}}
	\caption{Panels (a), (b), (c), and (d) follow the same format as in Figure~\ref{fig:ztrin0_1}, but correspond to a disk with a zero-torque boundary condition at a finite inner edge. The inner radius is set to $r_{\rm in} = 0.1 r_0$. Panels (e) and (f) show the bolometric luminosity, integrated from the inner edge to $r = 100r_0$, and its slope, $n_{\rm l} = \diff \ln L / \diff \ln t$, respectively. The dashed lines in panels (b), (d), and (f) represent a slope of $-(1 + l)$. The normalization constants $\dot{M}_0$ and $L_0$ are given by equation (\ref{eq:mdot0}) and equation (\ref{eq:lum0}), respectively.
	}
	\label{fig:ztrinf_1} 
\end{figure}
%

In Figure~\ref{fig:ztrinf_1}, we show the time evolution of the mass accretion rate, wind mass loss rate, and bolometric luminosity, computed by substituting equation~(\ref{eq:sigztf}) into equations~(\ref{eq:mdotacc}), (\ref{eq:mwind}), and (\ref{eq:bollum}), respectively. As in previous figures, the mass accretion and wind mass loss rates are normalized by $\dot{M}_0$ given by equation (\ref{eq:mdot0}), while the luminosity is normalized with $L_0$ given by equation (\ref{eq:lum0}). The inclusion of a magnetically driven wind enhances angular momentum transport, leading to faster inward motion of disk material and increased mass loss through the wind. This results in a more rapid decline in all three quantities—mass accretion rate, wind mass loss rate, and luminosity—compared to the case without wind. At late times, the temporal slopes of these quantities approach $-(1 + l)$, consistent with the asymptotic solutions given by equations (\ref{eq:mdotztf}) to (\ref{eq:lumztf}).  

%
\section{Case II: Zero mass accretion rate inner boundary condition} 
\label{sec:masszero}
%

In this section, we derive the Green's function and the corresponding surface density solution for a disk with a zero mass accretion rate at the inner edge, $r_{\rm in}$. A vanishing mass accretion rate at the inner edge implies that the total disk mass remains constant in the absence of wind, while it decreases over time in the presence of wind due to mass loss exclusively through the wind. Using equation (\ref{eq:sigsoln}) for the surface density and substituting it into equation (\ref{eq:mdotacc}) yields the mass accretion rate as
\begin{multline}\label{eq:mdoteqn}
	\dot{M} = 6\pi \nu_{\rm c} \int_{0}^{\infty} \left[C_1(k) r^{(1-\psi)/2}\frac{\partial}{\partial r}\left\{r^{(1+\psi)/4}J_{l}\left(\frac{k}{\beta}r^{\beta}\right)\right\}+C_2(k) r^{(1-\psi)/2}\frac{\partial}{\partial r}\left\{r^{(1+\psi)/4}Y_{l}\left(\frac{k}{\beta}r^{\beta}\right)\right\}\right] \\ e^{-3\nu_{\rm c} t k^2} \, \diff k.
\end{multline}

\subsection{Inner edge at the origin, $r_{\rm in} = 0$}

In the limit of $r \rightarrow 0$, $J_{l}\left(k r^{\beta} / \beta\right) \propto r^{\beta l}$ and $Y_{l}\left(k r^{\beta} / \beta\right) \propto r^{-\beta l}$, where $\beta l = \sqrt{(1+\psi)^2+4\psi/(\lambda-1)} / 4  > 0$. Using this, we estimate the terms with radial dependencies in equation (\ref{eq:mdoteqn}) as
\begin{eqnarray}
	r^{\frac{(1-\psi)}{2}}\frac{\partial}{\partial r}\left\{r^{\frac{(1+\psi)}{4}}J_{l}\left(\frac{k}{\beta}r^{\beta}\right)\right\} & \propto& \left[\frac{1+\psi+\sqrt{(1+\psi)^2+4\psi/(\lambda-1)}}{4}\right] r^{(\sqrt{(1+\psi)^2+4\psi/(\lambda-1)}-(1+\psi))/4} \nonumber\\
	\label{eq:Mdotc1}\\
	r^{\frac{(1-\psi)}{2}}\frac{\partial}{\partial r}\left\{r^{\frac{(1+\psi)}{4}}Y_{l}\left(\frac{k}{\beta}r^{\beta}\right)\right\} &\propto& \left[\frac{1+\psi-\sqrt{(1+\psi)^2+4\psi/(\lambda-1)}}{4}\right] r^{-(\sqrt{(1+\psi)^2+4\psi/(\lambda-1)}+1+\psi)/4} \nonumber \\
	\label{eq:Mdotc2}
\end{eqnarray}


When the wind is present, i.e., $\psi > 0$, equation (\ref{eq:Mdotc1}) approaches zero as $r_{\rm in} \to 0$, while equation (\ref{eq:Mdotc2}) diverges. Consequently, equation (\ref{eq:mdoteqn}) requires that the coefficient $C_2(k) = 0$ in order for the mass accretion rate to vanish at $r_{\rm in} = 0$. Substituting this condition into equation (\ref{eq:sigsoln}) yields a surface density profile identical to that given by equation (\ref{eq:sigsoln1}) for the case of a zero-torque boundary condition at $r_{\rm in} = 0$ (see Section~\ref{sec:zeror0}). This indicates that, in the presence of magnetically driven winds, the time-dependent surface density solution remains the same whether a zero-torque or zero-mass-accretion-rate boundary condition is imposed at the inner edge, provided $r_{\rm in} = 0$. This equivalence is further supported by the late-time solution for the mass accretion rate under the zero-torque boundary condition, given by equation~(\ref{eq:mdotztor}), where $\dot{M} \propto r^{\beta l - (1+\psi)/4}$. Since $\beta l - (1+\psi)/4 = \left[\sqrt{(1+\psi)^2 + 4\psi/(\lambda - 1)} - (1+\psi)\right]/4 > 0$ for $\psi > 0$, it follows that $\dot{M} \to 0$ as $r \to 0$.


\subsection{Inner edge at the finite radius, $r_{\rm in} > 0$}
\label{sec:zeromacccrinf}


In the case of a finite inner radius ($r_{\rm in} > 0$), applying the zero mass accretion rate condition at $r_{\rm in}$ to equation~(\ref{eq:mdoteqn}) yields \footnote{We have used the Bessel identities $\partial(z^l J_{l}(z))/\partial z = z^l J_{l-1}(z)$ and $\partial(z^l Y_{l}(z))/\partial z = z^l Y_{l-1}(z)$ for the derivation.}
\begin{equation}\label{eq:bndmfzero1}
	C_2(k) = - C_1(k) \left[\frac{\{(1+\psi)/4-\beta l\}J_{l}\left((k/\beta) r_{\rm in}^{\beta}\right) +k r_{\rm in}^{\beta}J_{l-1}\left((k/\beta) r_{\rm in}^{\beta}\right)}{\{(1+\psi)/4-\beta l\}Y_{l}\left((k/\beta) r_{\rm in}^{\beta}\right) +k r_{\rm in}^{\beta}Y_{l-1}\left((k/\beta) r_{\rm in}^{\beta}\right)} \right].
\end{equation}
Substituting equation (\ref{eq:bndmfzero1}) in equation (\ref{eq:sigsoln}) results in 
\begin{multline}\label{eq:sigsolmzerofn2}
	\frac{r_{\rm in}^{\beta}}{\beta}[r_{\rm in}x^{1/\beta}]^{(4n+1+\psi)/4}\Sigma(x,t) = \int_{0}^{\infty} \frac{C_1(k_1 \beta r_{\rm in}^{-\beta})\mathcal{Q}_{l}^2(k_1,\psi,n)}{\left[\left(\frac{1+\psi}{4}-\beta l\right)Y_{l}\left(k_1\right) + \beta k_1 Y_{l-1}\left(k_1\right)\right]k_1} \frac{\mathcal{W}_{l}(k_1,x,\psi,n)}{\mathcal{Q}_{l}^2(k_1,\psi,n)} k_1 \\ e^{-3\beta^2 \nu_{\rm c} r_{\rm in}^{-2\beta} t k_1^2}\, \diff k_1,
\end{multline}
where $k_1 = (k/\beta)r_{\rm in}^{\beta}$, $r = r_{\rm in}x^{1/\beta}$, and $\mathcal{W}_{l}(k_1,x,\psi,n)$ and $\mathcal{Q}_{l}^2(k_1,\psi,n)$ are given by
\begin{eqnarray}
	\mathcal{W}_{l}(k_1,x,\psi,n) &\equiv& W_{l}\left(k_1,x,\frac{1+\psi}{4},\beta\right)= J_{l}(k_1 x)\left[\left(\frac{1+\psi}{4}-\beta l\right)Y_{l}(k_1)+ \beta k_1  Y_{l-1}(k_1) \right] \nonumber\\
	&&- Y_{l}(k_1 x)\left[\left(\frac{1+\psi}{4}-\beta l\right)J_{l}(k_1)+ \beta k_1 J_{l-1}(k_1) \right] \label{eq:Wlt}\\ 
	\mathcal{Q}_{l}^2(k_1,\psi,n) &\equiv& Q_{l}^2\left(k_1,\frac{1+\psi}{4},\beta\right) =  \left[\left(\frac{1+\psi}{4}-\beta l\right)Y_{l}(k_1)+ \beta k_1 Y_{l-1}(k_1) \right]^2 \nonumber \\
	&&+\left[\left(\frac{1+\psi}{4}-\beta l\right)J_{l}(k_1)+ \beta k_1 J_{l-1}(k_1) \right]^2,\label{eq:Qlt}
\end{eqnarray}
estimated using equations (\ref{eq:Wlfn}) and (\ref{eq:Qlfn}), respectively. Applying the generalized Weber transform (equations \ref{eq:gweber} and \ref{eq:gweber1} in Appendix \ref{sec:inttrans}) to equation (\ref{eq:sigsolmzerofn2}) at $t = 0$ gives
\begin{equation}\label{eq:c1f}
	\frac{C_1(k_1 \beta r_{\rm in}^{-\beta})\mathcal{Q}_{l}^2(k_1,\psi,n)}{\left[\left(\frac{1+\psi}{4}-\beta l\right)Y_{l}\left(k_1\right) + \beta k_1 Y_{l-1}\left(k_1\right)\right]k_1} = \int_{1}^{\infty} \frac{r_{\rm in}^{\beta}}{\beta}[r_{\rm in}x^{'1/\beta}]^{(4n+1+\psi)/4}\Sigma(x^{'},t=0)  \mathcal{W}_{l}(k_1,x^{'},\psi,n) x^{'} \, \diff x^{'}.
\end{equation}
Substituting it in equation (\ref{eq:sigsolmzerofn2}), and using $r = r_{\rm in}x^{1/\beta}$, the surface density profile is given by
\begin{multline}\label{eq:sigsolmzerofn3}
	\Sigma(r,t) = \int_{r_{\rm in}}^{\infty} \Sigma(r^{'},t=0) r^{-(4n+1+\psi)/4}r^{'(4n+1+\psi)/4+2\beta-1}\frac{\beta}{r_{\rm in}^{2\beta}} \\ \int_{0}^{\infty} \frac{\mathcal{W}_{l}(k_1,(r^{'}/r_{\rm in})^{\beta},\psi,n)\mathcal{W}_{l}(k_1,(r/r_{\rm in})^{\beta},\psi,n)}{\mathcal{Q}_{l}^2(k_1,\psi,n)} e^{-3\beta^2 \nu_{\rm c} r_{\rm in}^{-2\beta} t k_1^2} k_1 \, \diff k_1 \diff r^{'}.
\end{multline}
By comparing equation (\ref{eq:sigsolmzerofn3}) with equation (\ref{eq:greenfncn}), the Green's function is 
\begin{multline}\label{eq:greensolmzerofn}
	\mathcal{G}(r,r^{'},t)=\frac{\beta}{r_{\rm in}}  \left(\frac{r}{r_{\rm in}}\right)^{-(4n+1+\psi)/4} \left(\frac{r^{'}}{r_{\rm in}}\right)^{(5+\psi)/4}\\ \int_{0}^{\infty} \frac{\mathcal{W}_{l}(k_1,(r^{'}/r_{\rm in})^{\beta},\psi,n)\mathcal{W}_{l}(k_1,(r/r_{\rm in})^{\beta},\psi,n)}{\mathcal{Q}_{l}^2(k_1,\psi,n)} e^{-3\beta^2 \nu_{\rm c} r_{\rm in}^{-2\beta} t k_1^2} k_1 \, \diff k_1 .
\end{multline}
Although an analytical solution for this Green's function is not available, equation (\ref{eq:greensolmzerofn}) provides its exact integral form. This expression can be used to numerically estimate the time evolution of the disk for any given initial surface density distribution.

\begin{figure}
	\centering
	\subfigure[]{\includegraphics[scale = 0.55]{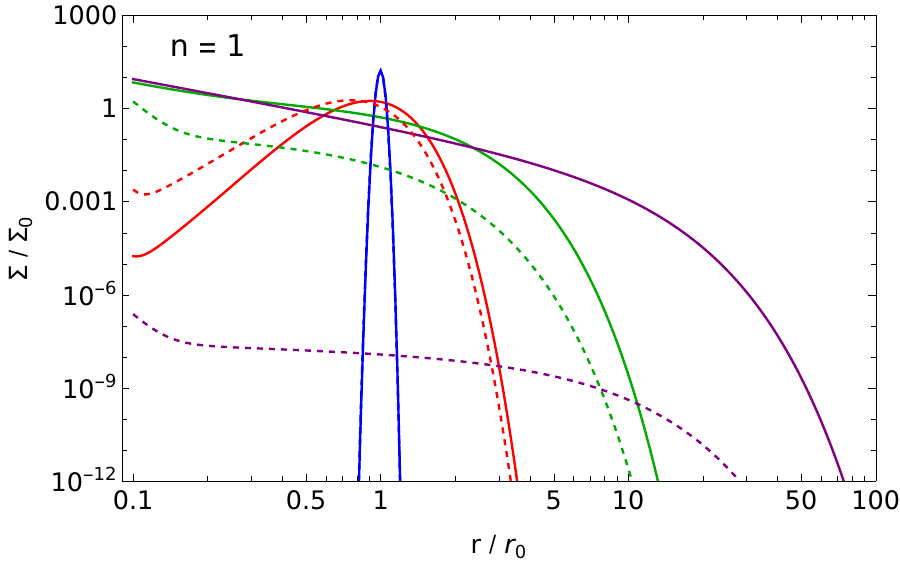}}
	\subfigure[]{\includegraphics[scale = 0.52]{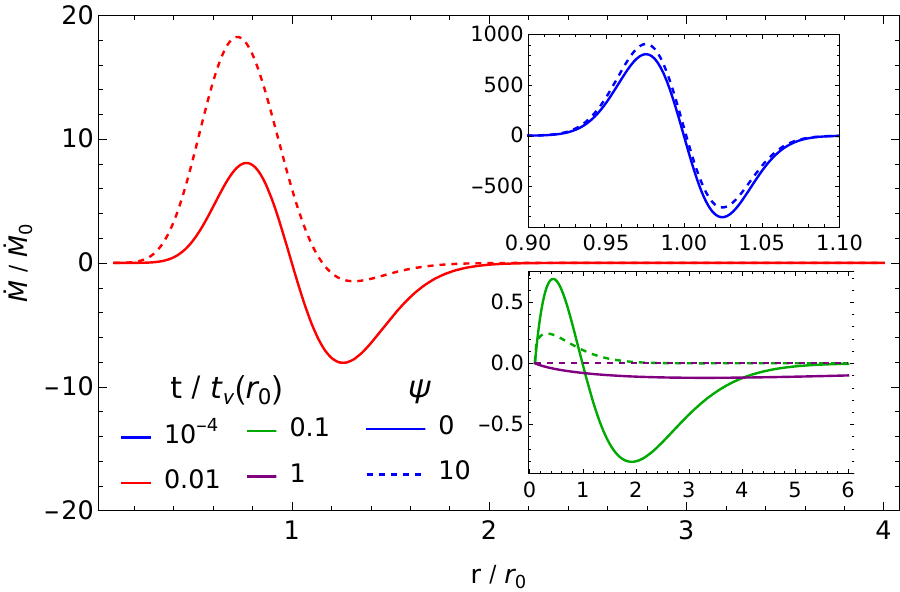}}
	\subfigure[]{\includegraphics[scale = 0.65]{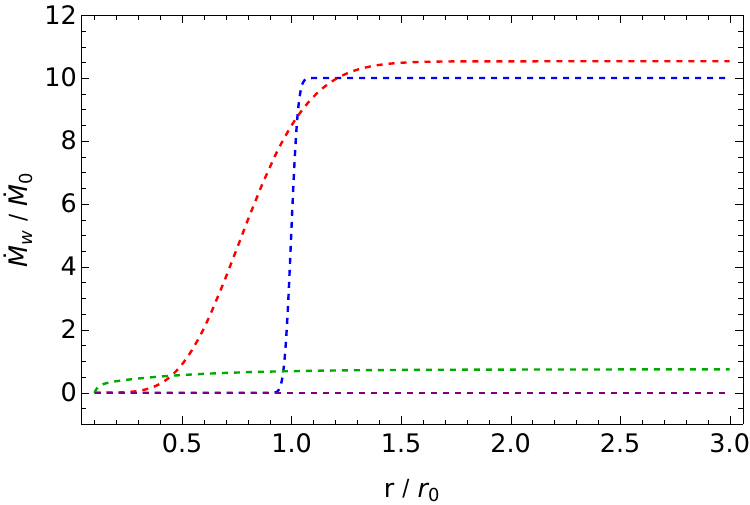}}
	\caption{
		The same format as in Figure~\ref{fig:ztrinf}, but for a disk with a zero mass accretion rate at a finite inner edge. The inner radius of the disk is set to $r_{\rm in} = 0.1 r_0$.
	}
	\label{fig:zmrinf} 
\end{figure}

The surface density using the initial surface density profile given by equation (\ref{eq:inisd}) and Green's function given by equation (\ref{eq:greensolmzerofn}) in (\ref{eq:greenfncn}) is given by
\begin{multline}\label{eq:sigsolmzerofns}
	\Sigma(r,t)= \Sigma_0 \beta \left(\frac{r}{r_{\rm in}}\right)^{-(4n+1+\psi)/4} \left(\frac{r_0}{r_{\rm in}}\right)^{(9+\psi)/4}\\ \int_{0}^{\infty} \frac{\mathcal{W}_{l}(k_1,(r_0/r_{\rm in})^{\beta},\psi,n)\mathcal{W}_{l}(k_1,(r/r_{\rm in})^{\beta},\psi,n)}{\mathcal{Q}_{l}^2(k_1,\psi,n)} e^{-3\beta^2 (r_0 /r_{\rm in})^{2\beta} (t / t_{\nu}(r_0)) k_1^2} k_1 \, \diff k_1 .
\end{multline}
Figure~\ref{fig:zmrinf} illustrates the radial evolution of the surface density $\Sigma$, the mass accretion rate $\dot{M}$, and the outwardly integrated wind mass loss rate $\dot{M}_{\rm w}$ for disks without wind ($\psi = 0$, solid lines) and with wind ($\psi = 10$, dashed lines). Panel (a) shows that the surface density is finite at the inner edge, $r_{\rm in} = 0.1r_0$, while panel (b) confirms that the mass accretion rate vanishes at the inner edge, consistent with the boundary condition of zero mass flux.

%

To derive the late-time behavior of the disk, where $t > t_{\nu}(r_0) > t_{\nu}(r_{\rm in})$, we note that $(r_0/r_{\rm in})^{2\beta},(t/t_{\nu}(r_0)) = t/t_{\nu}(r_{\rm in}) > 1$, in which case only modes with $k_1^2 \ll 1$ are relevant. When the wind is absent ($\psi = 0$), an analytical late-time solution can be obtained for the inner disk region when $0 < l < 1$ (corresponding to $n < 3/2$, from equation~(\ref{eq:l})). In the limits $k_1 \ll 1$ and $k_1 (r_0/r_{\rm in})^{\beta} \ll 1$, we find $\mathcal{W}_{l}(k_1,(r/r{\rm in})^{\beta},\psi,n) \approx (2\beta/\pi)(r/r_{\rm in})^{-1/4}$ and $\mathcal{Q}_{l}^2(k_1,\psi,n) \approx 2^{2-2l}\csc^2(l\pi)[\Gamma(l)]^{-2}\beta^2 k_1^{2l}$, retaining only the leading terms. Substituting these into equation~(\ref{eq:sigsolmzerofns}) yields the late-time surface density for $r < r_0$ as
\begin{equation}\label{eq:sigasympzmacc}
	\Sigma(r<r_0, t> t_{\nu}(r_0)) = \frac{2^{2l-1}3^{-1+l} \beta^{-1+2l} \Gamma(l)}{\pi \csc(l\pi)} \Sigma_0 \left(\frac{r}{r_0}\right)^{-n-1/2} \left(\frac{t}{t_{\nu}(r_0)}\right)^{-1+l}.
\end{equation} 
We can see from equation (\ref{eq:mdotacc}) that this surface density will result in a zero mass accretion rate at the inner edge of the disk.

When the wind is present, $\psi > 0$, we obtain $\mathcal{W}_{l}(k_1,(r/r_{\rm in})^{\beta},\psi,n) \approx \left[\beta l-(1+\psi)/4\right](l \pi)^{-1} (r/r_{\rm in})^{\beta l} + \left[\beta l+(1+\psi)/4\right](l \pi)^{-1} (r/r_{\rm in})^{-\beta l}$ and $\mathcal{Q}_{l}^2(k_1,\psi,n) \approx \left[\beta l-(1+\psi)/4\right]^2 [\Gamma(l)/\pi]^2 (k_1/2)^{-2l}$. By substituting these in equation (\ref{eq:sigsolmzerofns}) gives the late time evolution of the surface density near the inner region of the disk ($r < r_0$) as
\begin{multline}
	\Sigma(r<r_0, t> t_{\nu}(r_0)) = \frac{3^{-1-l} 2^{-1-2l}\beta^{-1-2l} \Sigma_0}{[(1+\psi)/4-\beta l]^2 \Gamma(1+l)} \left[\beta l -\frac{(1+\psi)}{4}+\left(\beta l + \frac{1+\psi}{4}\right) \left(\frac{r_{\rm in}}{r_0}\right)^{2\beta l}\right]\\ \left(\frac{r}{r_{0}}\right)^{-n+\beta l-(1+\psi)/4} \left[\beta l -\frac{(1+\psi)}{4}+\left(\beta l + \frac{1+\psi}{4}\right) \left(\frac{r_{\rm in}}{r}\right)^{2\beta l}\right] \left(\frac{t}{t_{\nu}(r_0)}\right)^{-(1+l)}.
\end{multline}
It can be seen that the surface density is non-zero at the inner edge of the disk indicating a finite torque at the inner boundary. By substituting this surface density in equation (\ref{eq:mdotacc}), the mass accretion rate is
\begin{multline}
	\dot{M}(r<r_0, t> t_{\nu}(r_0)) = \frac{4 \beta^{-1-2 l}\dot{M}_0 }{12^{1+l} \Gamma(1+l)} \left(\frac{4\beta l + 1+\psi}{4\beta l - 1-\psi}\right)\left[\beta l -\frac{(1+\psi)}{4}+\left(\beta l + \frac{1+\psi}{4}\right) \left(\frac{r_{\rm in}}{r_0}\right)^{2\beta l}\right]\\ \left(\frac{r}{r_0}\right)^{\beta l - (1+\psi)/4} \left[1-\left(\frac{r_{\rm in}}{r}\right)^{2\beta l}\right]\left(\frac{t}{t_{\nu}(r_0)}\right)^{-(1+l)}.
\end{multline}
The mass accretion rate is zero at the inner edge of the disk, thus satisfying the inner boundary condition. For $r \gg r_{\rm in}$, the radial evolution of the mass accretion rate is the same as in the case with a zero-torque boundary condition (see Equation~\ref{eq:mdotztf}). The integrated wind mass loss rate using equation (\ref{eq:mwind}) is 
\begin{multline}\label{eq:mwasymzmacc}
	\dot{M}_{\rm w}(r<r_0, t> t_{\nu}(r_0)) = \frac{\psi \beta^{-1-2l}}{3^{1+l} 2^{2+2l} (\lambda-1)} \frac{\dot{M}_0}{[(1+\psi)/4-\beta l]^2 \Gamma(1+l)} \left[\beta l -\frac{(1+\psi)}{4} \right. \\ \left. +\left(\beta l + \frac{1+\psi}{4}\right) \left(\frac{r_{\rm in}}{r_0}\right)^{2\beta l}\right] \left(\frac{r}{r_{0}}\right)^{\beta l - (1+\psi)/4} \left[1-\left(\frac{r_{\rm in}}{r}\right)^{(1+\psi)/2}\right] \left(\frac{t}{t_{\nu}(r_0)}\right)^{-(1+l)},
\end{multline}
and follows the same time profiles as mass accretion rate in the inner region of the disk. The bolometric luminosity derived using equation (\ref{eq:bollum}) is
\begin{multline}\label{eq:lumasymzmacc}
	L(r<r_0, t> t_{\nu}(r_0)) = \frac{3^{-1-l} 2^{-1-2l}\beta^{-1-2l}L_0}{ \left[(1+\psi)/4-\beta l\right]^2 \Gamma(1+l)} \left[3+\frac{\psi(2\lambda-3)}{2(\lambda-1)}\right] \left[\beta l -\frac{(1+\psi)}{4} \right. \\ \left. +\left(\beta l + \frac{1+\psi}{4}\right) \left(\frac{r_{\rm in}}{r_0}\right)^{2\beta l}\right] \left(\frac{r_{\rm t}}{r_{0}}\right)^{\beta l - (5+\psi)/4}\left[\frac{4\beta l-(1+\psi)}{4\beta l-(5+\psi)} -\frac{32 \beta l}{16 \beta^2l^2-(5+\psi)^2} \left(\frac{r_{\rm in}}{r_t}\right)^{\beta l -(5+\psi)/4} \right. \\ \left.- \left(\frac{4\beta l+1+\psi}{4\beta l+5+\psi}\right)\left(\frac{r_{\rm in}}{r_t}\right)^{2\beta l}\right] \left(\frac{t}{t_{\nu}(r_0)}\right)^{-(1+l)},
\end{multline}
where $r_t$ is the truncation radius below $r_0$, inside which the disk has had sufficient time to approach the asymptotic solution. It can be seen that the mass accretion rate, wind mass loss rate and the bolometric luminosity in the inner region of the disk follows a time profile of $t^{-(1+l)}$, which is same as in the case of zero torque at the finite $r_{\rm in}$.

%
\begin{figure}
	\centering
	\subfigure[]{\includegraphics[scale = 0.63]{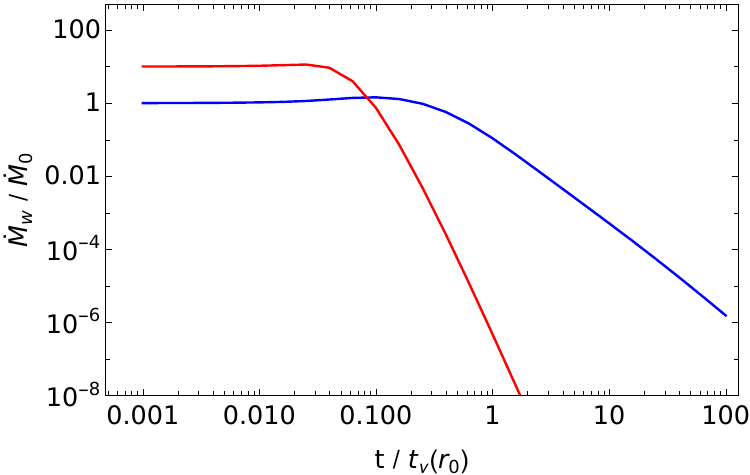}}
	\subfigure[]{\includegraphics[scale = 0.61]{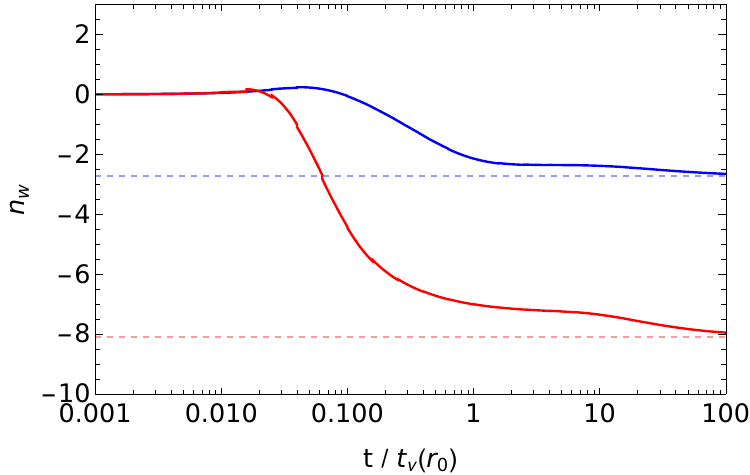}}
	\subfigure[]{\includegraphics[scale = 0.63]{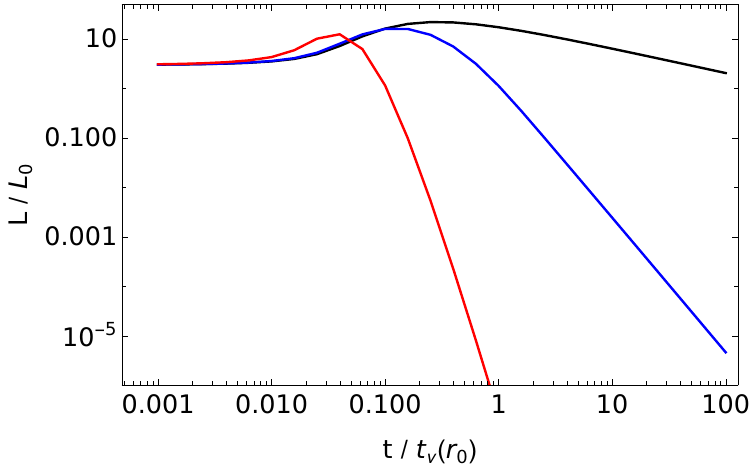}}
	\subfigure[]{\includegraphics[scale = 0.61]{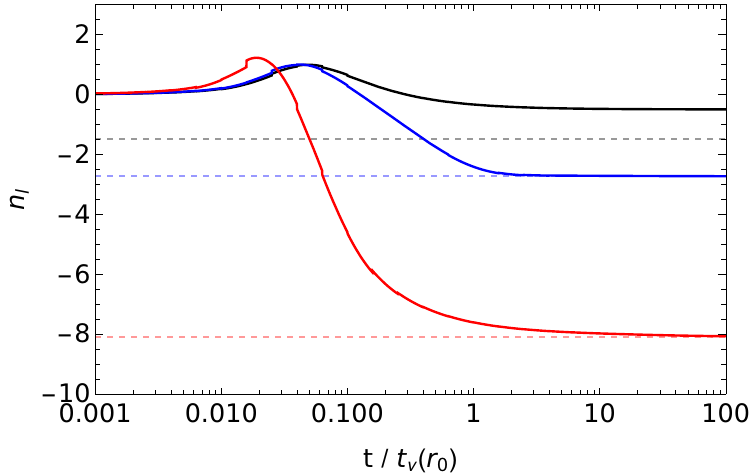}}
	\caption{The same format as in Figure~\ref{fig:ztrinf_1}, but correspond to a disk with a zero-mass accretion rate boundary condition at a finite inner edge. The inner radius is set to $r_{\rm in} = 0.1r_0$. The dashed lines in panels (b) and (d) represent a slope of $-(1 + l)$.
	}
	\label{fig:zmrinf_1} 
\end{figure}
%

In Figure~\ref{fig:zmrinf_1}, we show the time evolution of the wind mass loss rate and the bolometric luminosity, computed by substituting equation~(\ref{eq:sigsolmzerofns}) into equations~(\ref{eq:mwind}) and (\ref{eq:bollum}), respectively. Since the mass accretion rate is zero at the inner boundary, we do not show its evolution here. At late times, the slopes of both the wind mass loss rate and the luminosity asymptotically approach $-(1 + l)$, consistent with equations~(\ref{eq:mwasymzmacc}) and (\ref{eq:lumasymzmacc}). However, in the absence of a wind, $\psi = 0$, the slope of the luminosity differs and instead follows $-1 + l$. This is because the luminosity is the radial integral of the surface density as reflected in equation~(\ref{eq:bollum}), and the late-time behavior of the surface density in the absence of wind follows $t^{-1 + l}$, as seen from equation (\ref{eq:sigasympzmacc}). Consequently, the luminosity also evolves as $t^{-1 + l}$ in that case. In contrast, for a disk with a zero-torque boundary condition at a finite inner edge (see Figure~\ref{fig:ztrinf_1} in Section~\ref{sec:ztorrinf}), the luminosity at late times follows $t^{-(1 + l)}$ even in the absence of wind. Therefore, when the wind is present, the luminosity exhibits the same temporal evolution for both boundary conditions. However, in the absence of wind, the luminosity declines more slowly for the zero mass accretion rate boundary condition compared to the zero-torque case.

%
\section{Case III: Finite torque and the mass accretion rate at the inner boundary} 
\label{sec:SFrin}
%

From equation~(\ref{eq:torque1}), the condition for a finite torque at the inner radius $r_{\rm in}$ in the limit $r_{\rm in} \rightarrow 0$ leads to
\begin{equation}
	C_1(k) r_{\rm in}^{\left(1-\psi + \sqrt{(1+\psi)^2+4\psi/(\lambda-1)}\right)/4} + C_2(k) r_{\rm in}^{\left(1-\psi - \sqrt{(1+\psi)^2+4\psi/(\lambda-1)}\right)/4} \neq 0,
\end{equation}
where we have used $J_{l}\left((k/\beta), r^{\beta}\right) \propto r^{\beta l}$ and $Y_{l}\left((k/\beta), r^{\beta}\right) \propto r^{-\beta l}$ as $r \rightarrow 0$ in equation~(\ref{eq:torque1}). In the limit of $r_{\rm in} = 0$, the second term diverges while the first term vanishes for $\psi > 0$ . Therefore, maintaining a finite torque at $r_{\rm in} = 0$ requires $C_2(k) \rightarrow 0$. This implies that the contribution from the $C_2(k)$ component to the surface density (equation \ref{eq:sigsoln}) is negligible for the disk evolution, and a non-zero $C_1(k)$ corresponds to a solution with zero torque at the center. Thus, it appears that a disk with finite torque at $r_{\rm in} = 0$ may not be physically realizable when the wind is present.

For the case of $r_{\rm in}>0$, we utilize the mass accretion rate equation to estimate the boundary condition. The mass accretion rate given by equation (\ref{eq:mdotacc}) results in $\dot{M} = 3 \pi (1+\psi) \nu_{\rm c} r^{n}\Sigma + 6 \pi \nu_{\rm c} r \partial (r^n \Sigma)/\partial r$. Equation~(\ref{eq:torque1}) indicates that a finite torque at $r_{\rm in} > 0$ has $\Sigma (r_{\rm in}) \neq 0$. Therefore, we impose the condition of 
\begin{equation}\label{eq:smfbnd}
	\frac{\partial}{\partial r}(r^n \Sigma) = 0
\end{equation}
at the inner edge $r_{\rm in}$, which results in a finite mass accretion rate at the $r_{\rm in}$ given by $\dot{M} = 3 \pi (1+\psi) \nu_{\rm c} r^{n}\Sigma$. By substituting surface density from equation (\ref{eq:sigsoln}) in equation (\ref{eq:smfbnd}) results in the following relation:
\begin{equation}
		C_2(k) = - C_1(k) \left[\frac{-\left\{\frac{1+\psi}{4}+\beta l\right\}J_{l}\left(\frac{k}{\beta}r_{\rm in}^{\beta}\right)+ k r_{\rm in}^{\beta} J_{l-1}\left(\frac{k}{\beta}r_{\rm in}^{\beta}\right)}{-\left\{\frac{1+\psi}{4}+\beta l\right\}Y_{l}\left(\frac{k}{\beta}r_{\rm in}^{\beta}\right)+ k r_{\rm in}^{\beta} Y_{l-1}\left(\frac{k}{\beta}r_{\rm in}^{\beta}\right)}\right].
\end{equation}
Substituting this in equation (\ref{eq:sigsoln}) and following the procedure similar to that in section \ref{sec:zeromacccrinf}, we get 
\begin{multline}\label{eq:smfinrin}
	\Sigma(r,t)= \int_{r_{\rm in}}^{\infty} \Sigma(r^{'},t=0) \frac{\beta}{r_{\rm in}}\left(\frac{r}{r_{\rm in}}\right)^{-(4n+1+\psi)/4} \left(\frac{r^{'}}{r_{\rm in}}\right)^{(4n+1+\psi)/4+2\beta-1} \\ \int_{0}^{\infty} \frac{\mathcal{W}_{l,1}(k_1,(r^{'}/r_{\rm in})^{\beta},\psi,n)\mathcal{W}_{l,1}(k_1,(r/r_{\rm in})^{\beta},\psi,n)}{\mathcal{Q}_{l,1}^2(k_1,\psi,n)} e^{-3\beta^2 \nu_{\rm c} r_{\rm in}^{-2\beta} t k_1^2} k_1 \, \diff k_1 \diff r^{'},
\end{multline}
where $k_1 = (k / \beta) r_{\rm in}^{\beta}$ and 
\begin{eqnarray}
	\mathcal{W}_{l,1}(k_1,x,\psi,n) &\equiv& W_{l}\left(k_1,x,-\frac{1+\psi}{4},\beta\right)= J_{l}(k_1 x)\left[-\left(\frac{1+\psi}{4}+\beta l\right)Y_{l}(k_1)+ \beta k_1  Y_{l-1}(k_1) \right] \nonumber\\
	&&- Y_{l}(k_1 x)\left[-\left(\frac{1+\psi}{4}+\beta l\right)J_{l}(k_1)+ \beta k_1 J_{l-1}(k_1) \right] \label{eq:Wlt1}\\ 
	\mathcal{Q}_{l,1}^2(k_1,\psi,n) &\equiv& Q_{l}^2\left(k_1,-\frac{1+\psi}{4},\beta\right) =  \left[-\left(\frac{1+\psi}{4}+\beta l\right)Y_{l}(k_1)+ \beta k_1 Y_{l-1}(k_1) \right]^2 \nonumber \\
	&&+\left[-\left(\frac{1+\psi}{4}+\beta l\right)J_{l}(k_1)+ \beta k_1 J_{l-1}(k_1) \right]^2.\label{eq:Qlt1}
\end{eqnarray}
By comparing equation (\ref{eq:smfinrin}) with equation (\ref{eq:greenfncn}), the Green's function is 
\begin{multline}\label{eq:greensmfinrin}
	\mathcal{G}(r,r^{'},t)=\frac{\beta}{r_{\rm in}}\left(\frac{r}{r_{\rm in}}\right)^{-(4n+1+\psi)/4} \left(\frac{r^{'}}{r_{\rm in}}\right)^{(5+\psi)/4} \\ \int_{0}^{\infty} \frac{\mathcal{W}_{l,1}(k_1,(r^{'}/r_{\rm in})^{\beta},\psi,n)\mathcal{W}_{l,1}(k_1,(r/r_{\rm in})^{\beta},\psi,n)}{\mathcal{Q}_{l,1}^2(k_1,\psi,n)} e^{-3\beta^2 \nu_{\rm c} r_{\rm in}^{-2\beta} t k_1^2} k_1 \, \diff k_1 .
\end{multline}

\begin{figure}
	\centering
	\subfigure[]{\includegraphics[scale = 0.54]{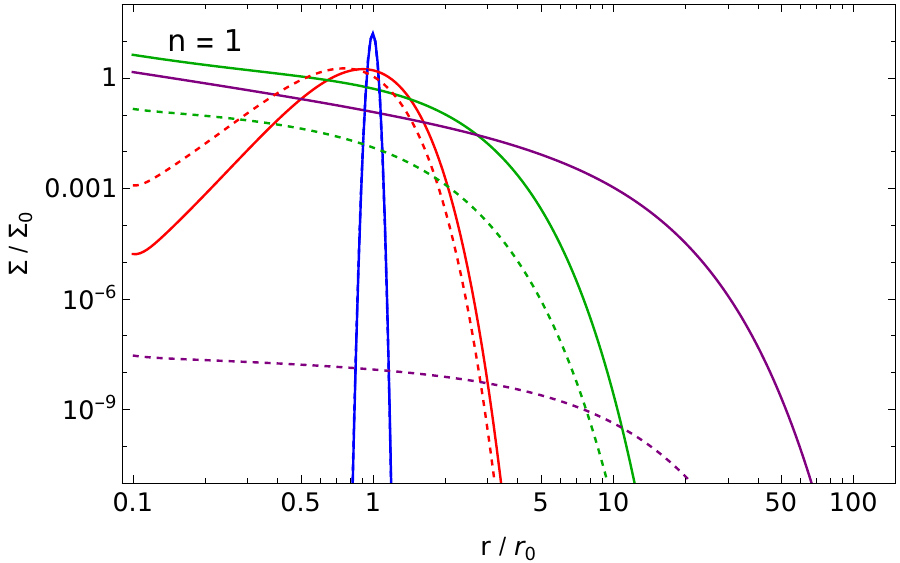}}
	\subfigure[]{\includegraphics[scale = 0.53]{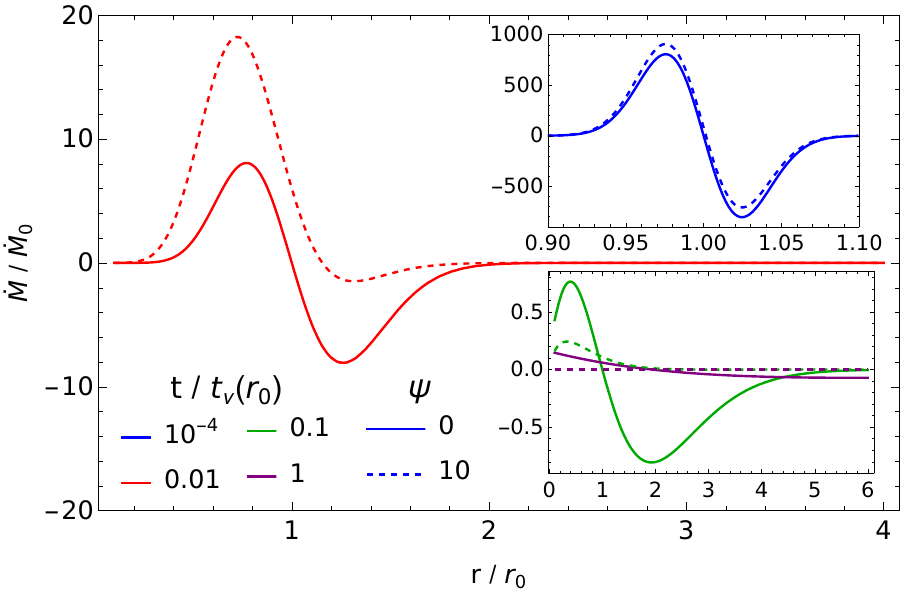}}
	\subfigure[]{\includegraphics[scale = 0.57]{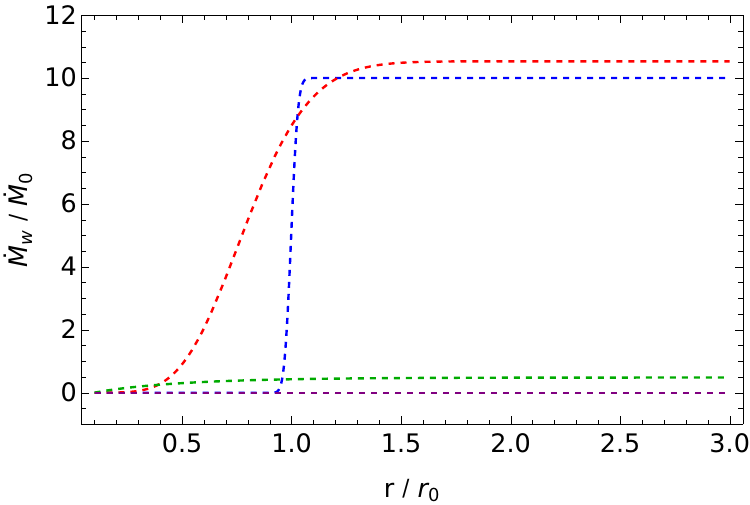}}
	\caption{
		The same format as in Figure~\ref{fig:ztrinf}, but for a disk with finite torque and mass accretion rate at a finite inner edge. The inner radius of the disk is set to $r_{\rm in} = 0.1 r_0$.
	}
	\label{fig:finsigmass} 
\end{figure}

The surface density, obtained by substituting the initial surface density profile from equation~(\ref{eq:inisd}) and the Green’s function from equation~(\ref{eq:greensmfinrin}) into equation~(\ref{eq:greenfncn}), is given by
\begin{multline}\label{eq:smfinrin}
	\Sigma(r,t)= \Sigma_0 \beta \left(\frac{r}{r_{\rm in}}\right)^{-(4n+1+\psi)/4} \left(\frac{r_0}{r_{\rm in}}\right)^{(9+\psi)/4}\\ \int_{0}^{\infty} \frac{\mathcal{W}_{l,1}(k_1,(r_0/r_{\rm in})^{\beta},\psi,n)\mathcal{W}_{l,1}(k_1,(r/r_{\rm in})^{\beta},\psi,n)}{\mathcal{Q}_{l,1}^2(k_1,\psi,n)} e^{-3\beta^2 (r_0 /r_{\rm in})^{2\beta} (t / t_{\nu}(r_0)) k_1^2} k_1 \, \diff k_1 .
\end{multline}
Figure~\ref{fig:finsigmass} illustrates the radial profiles of the surface density $\Sigma$, the mass accretion rate $\dot{M}$, and the outwardly integrated wind mass loss rate $\dot{M}_{\rm w}$ for disks without wind ($\psi = 0$, solid lines) and with wind ($\psi = 10$, dashed lines). Panels (a) and (b) show that both the surface density and mass accretion rate remain finite at the inner edge, $r_{\rm in} = 0.1r_0$, indicating a finite torque and mass accretion rate at the inner edge.

%
%

%
\begin{figure}
	\centering
	\subfigure[]{\includegraphics[scale = 0.63]{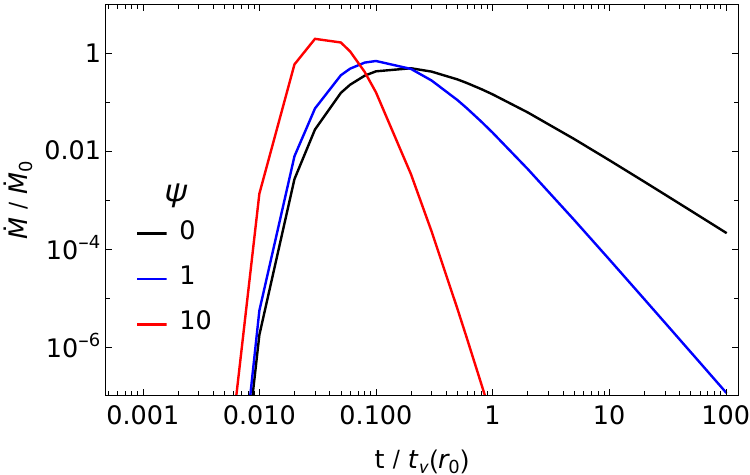}}
	\subfigure[]{\includegraphics[scale = 0.61]{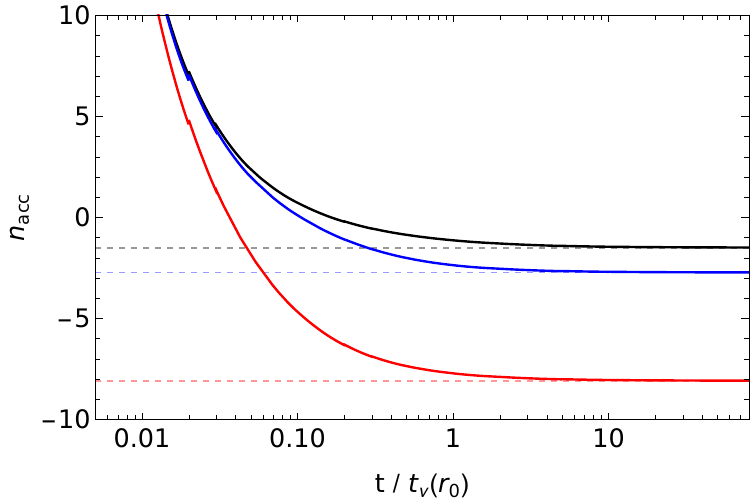}}
	\subfigure[]{\includegraphics[scale = 0.63]{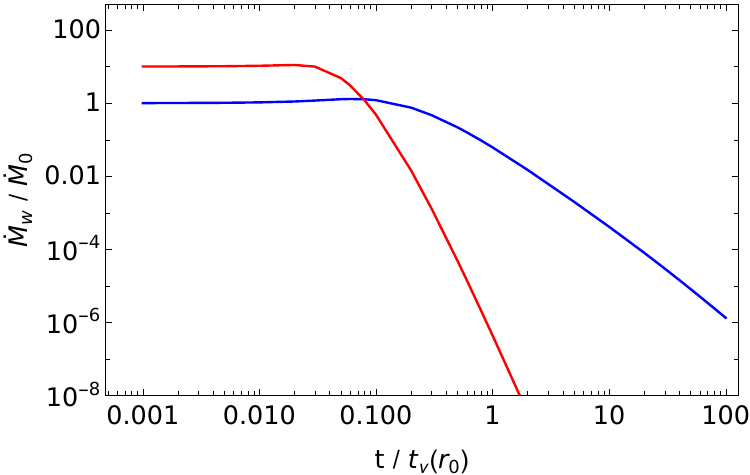}}
	\subfigure[]{\includegraphics[scale = 0.61]{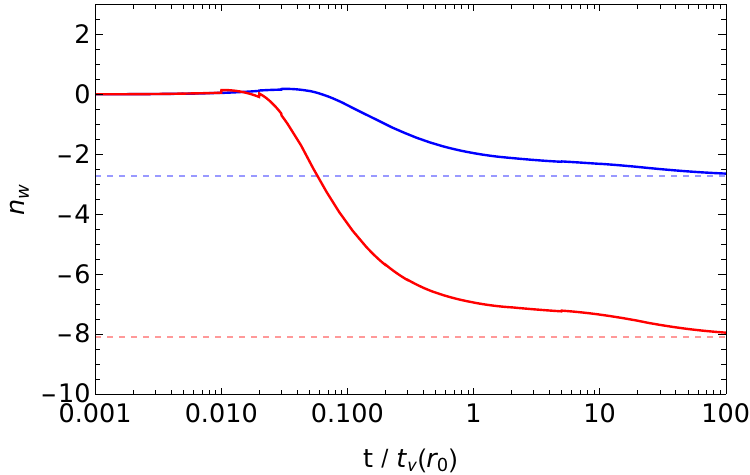}}
	\subfigure[]{\includegraphics[scale = 0.63]{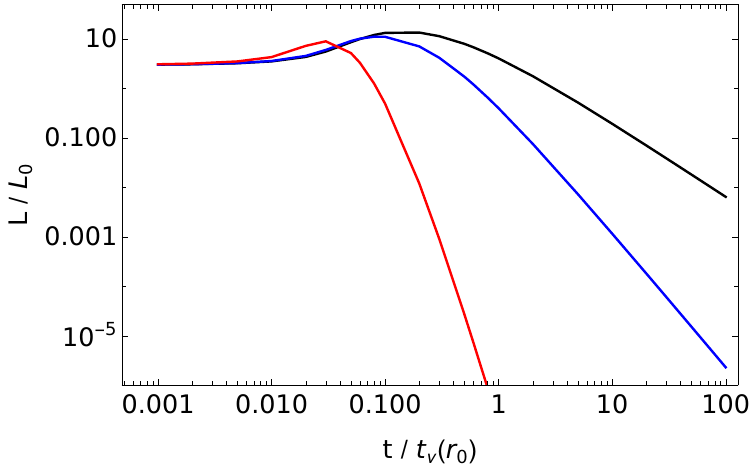}}
	\subfigure[]{\includegraphics[scale = 0.61]{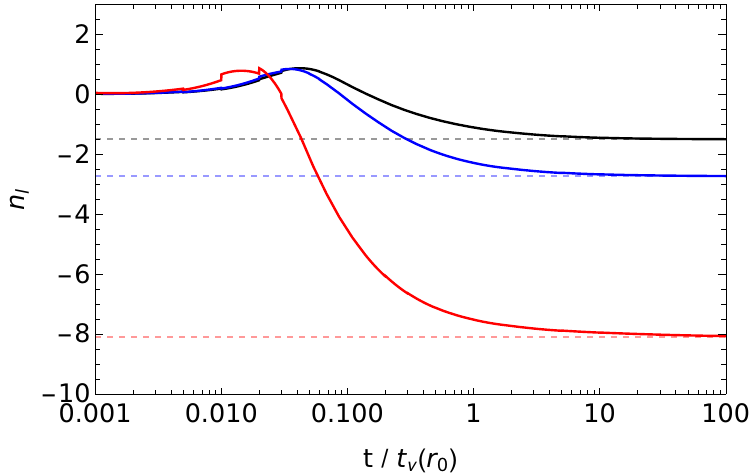}}
	\caption{The same format as in Figure~\ref{fig:ztrinf_1}, but correspond to a disk with finite torque and mass accretion rate boundary condition at the finite inner edge. The inner radius is set to $r_{\rm in} = 0.1 r_0$. The dashed lines in panels (b), (d) and (e) represent a slope of $-(1 + l)$.
	}
	\label{fig:finsigmass_1} 
\end{figure}
%

The late time behaviour of the disk is derived following the analysis similar to that in section \ref{sec:zeromacccrinf}, and the resultant surface density is given by
\begin{multline}
\Sigma(r<r_0, t>t_{\nu}(r_0)) = \frac{3^{-1-l}2^{-1-2l}\beta^{-1-2l}\Sigma_0}{\left[(1+\psi)/4+\beta l\right]^2 \Gamma(1+l)} \left[\beta l +\frac{1+\psi}{4} +\left(\beta l - \frac{1+\psi}{4}\right)\left(\frac{r_{\rm in}}{r_0}\right)^{2\beta l}\right]\\ \left(\frac{r}{r_{0}}\right)^{-n+\beta l -(1+\psi)/4} \left[\beta l + \frac{1+\psi}{4}+ \left(\beta l - \frac{1+\psi}{4}\right)\left(\frac{r_{\rm in}}{r}\right)^{2\beta l}\right] \left(\frac{t}{t_{\nu}(r_0)}\right)^{-(1+l)}.
\end{multline}
The temporal evolution of the surface density at late times is identical to the cases of zero torque and zero mass accretion rate at the inner edge. Similarly, we estimate the late time evolution of the mass accretion rate and the mass loss rate to be 
\begin{eqnarray}
\dot{M}=&& \frac{3^{-1-l}2^{-2l}\pi \beta^{-1-2l} \dot{M}_0}{\left[(1+\psi)/4 + \beta l\right]^2 \Gamma(1+l)} \left[\beta l +\frac{1+\psi}{4}+\left(\beta l - \frac{1+\psi}{4}\right)\left(\frac{r_{\rm in}}{r_0}\right)^{2\beta l}\right] \left(\frac{r}{r_{0}}\right)^{\beta l - (1+\psi)/4} \nonumber \\
&&\left[\left(\beta l +\frac{1+\psi}{4}\right)^2-\left(\beta l-\frac{1+\psi}{4}\right)^2\left(\frac{r_{\rm in}}{r}\right)^{2\beta l}\right]\left(\frac{t}{t_{\nu}(r_0)}\right)^{-(1+l)}, \\
\dot{M}_{\rm w}=&& \frac{3^{-1-l}2^{-2l}\beta^{-1-2l} \dot{M}_0}{\Gamma(1+l)} \left[\beta l +\frac{1+\psi}{4}+\left(\beta l - \frac{1+\psi}{4}\right)\left(\frac{r_{\rm in}}{r_0}\right)^{2\beta l}\right] \left(\frac{r}{r_{0}}\right)^{\beta l -(1+\psi)/4}\nonumber \\
&&  \left[1-\frac{\beta l (1+\psi)}{[\beta l + (1+\psi)/4]^2} \left(\frac{r_{\rm in}}{r}\right)^{\beta l- (1+\psi)/4} - \left(\frac{4\beta l - (1+\psi)}{4\beta l + 1+ \psi}\right)^2 \left(\frac{r_{\rm in}}{r}\right)^{2\beta l}\right] \left(\frac{t}{t_{\nu}(r_0)}\right)^{-(1+l)}.
\end{eqnarray}
It can be seen that the mass accretion rate is finite at the inner edge, and both mass accretion and loss rates follows the same power-law decay with time in the inner regions of the disk ($r < r_0$). The late time evolution of the bolometric luminosity is 
\begin{multline}
L = \frac{3^{-1-l}2^{-1-2l}\beta^{-1-2l}L_0}{[\beta l +(1+\psi)/4]^2 \Gamma(1+l)} \left[3+\frac{(2\lambda-3)\psi}{2(\lambda-1)}\right]\left[\beta l +\frac{1+\psi}{4} +\left(\beta l - \frac{1+\psi}{4}\right)\left(\frac{r_{\rm in}}{r_0}\right)^{2\beta l}\right] \\ \left(\frac{r_t}{r_{0}}\right)^{\beta l -(5+\psi)/4} \left[\frac{4\beta l +1+\psi}{4\beta l - (5+\psi)}  - \frac{16\beta l (3+\psi)}{16\beta^2 l^2 -(5+\psi)^2}\left(\frac{r_{\rm in}}{r_t}\right)^{\beta l - (5+\psi)/4}-\frac{4\beta l - (1+\psi)}{4\beta l +5+\psi} \left(\frac{r_{\rm in}}{r_t}\right)^{2\beta  l}\right]  \\ \left(\frac{t}{t_{\nu}(r_0)}\right)^{-(1+l)},
\end{multline}
where $r_t$ is the truncation radius below $r_0$, inside which the disc has had sufficient time to approach the asymptotic solution.

In Figure~\ref{fig:finsigmass_1}, we show the time evolution of the mass accretion rate, wind mass loss rate, and bolometric luminosity, computed numerically by substituting equation~(\ref{eq:smfinrin}) into equations~(\ref{eq:mdotacc}), (\ref{eq:mwind}), and (\ref{eq:bollum}), respectively. As in previous figures, the mass accretion and wind mass loss rates are normalized by $\dot{M}_0$ given by equation (\ref{eq:mdot0}), while the luminosity is normalized with $L_0$ given by equation (\ref{eq:lum0}). The presence of magnetically driven wind results in a more rapid decline in all three quantities—mass accretion rate, wind mass loss rate, and luminosity—compared to the case without wind. At late times, the temporal slopes of these quantities approach $-(1 + l)$, consistent with the asymptotic solutions estimated above.

%
\section{Comparison of Disk Physical Quantities under Three Boundary Conditions} 
\label{sec:comp_par}
%

In this section, we compare the disk mass, mass accretion rate, and wind mass loss rate estimated for the three cases . Case I represents a disk with zero torque at the inner edge, Case II corresponds to a disk with zero mass accretion rate at the inner edge, and Case III describes a disk with finite torque and finite mass accretion rate at the inner edge. 

\subsection{Comparison of zero torque (case I) solution of $r_{\rm in} \rightarrow 0$ and $r_{\rm in} > 0$}

%
\begin{figure}
	\centering
	\subfigure[]{\includegraphics[scale = 0.63]{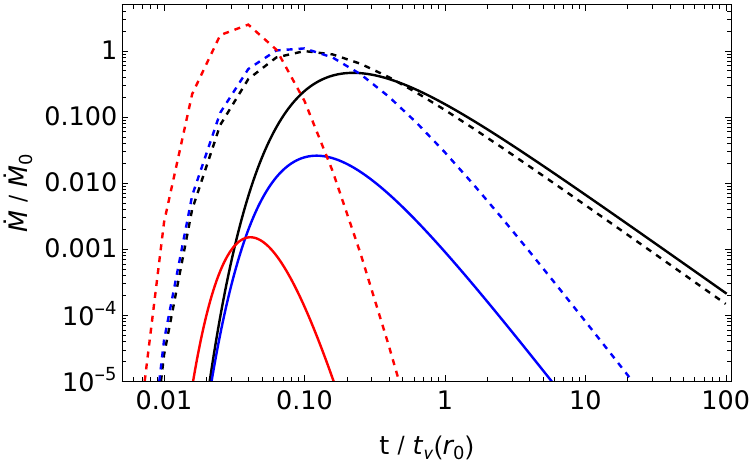}}
	\subfigure[]{\includegraphics[scale = 0.63]{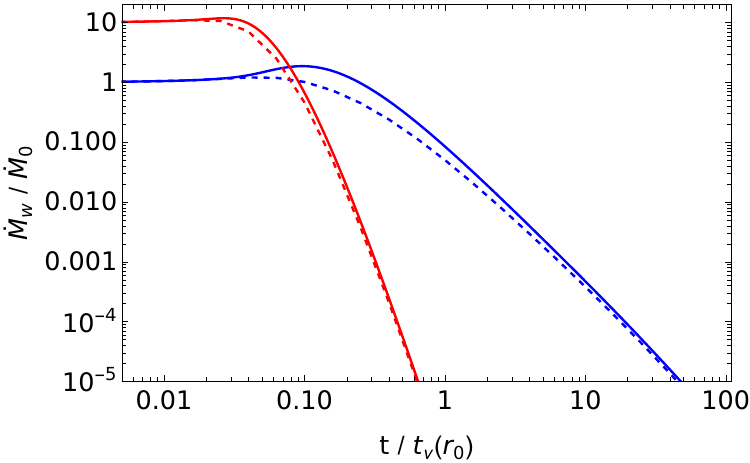}}
	\subfigure[]{\includegraphics[scale = 0.63]{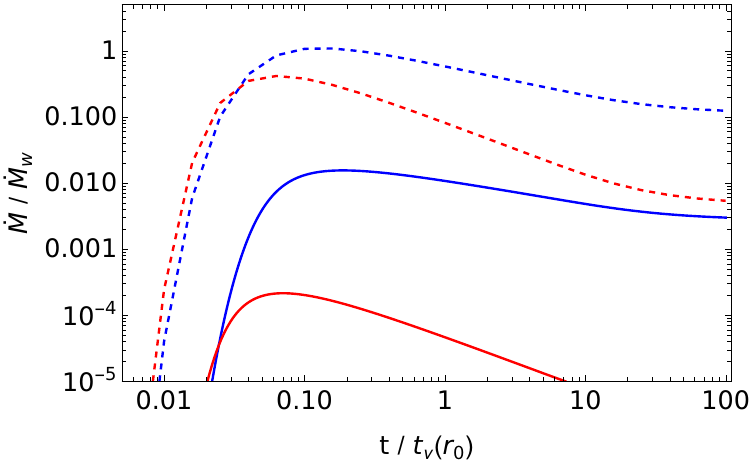}}
	\subfigure[]{\includegraphics[scale = 0.63]{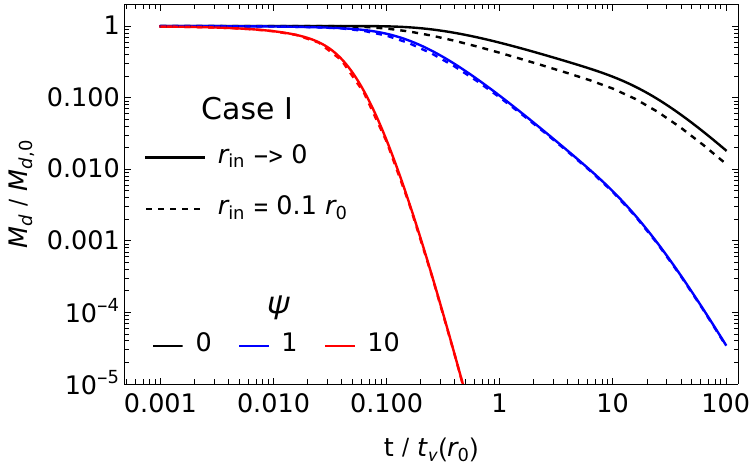}}
	\caption{Comparison of the time evolution of the mass accretion rate (panel a), wind mass-loss rate (panel b), the ratio of mass accretion to wind-loss rates (panel c), and disk mass (panel d) for Case I (zero-torque condition), comparing disks with different inner radii. Different colored lines represent different values of $\psi$, while solid and dashed lines correspond to disks extending to the origin and at a finite inner radius, respectively. The black line is absent in panels (b) and (c) because $\psi = 0$ corresponds to a disk without wind.}
	\label{fig:phycomp_caseI} 
\end{figure}
%

The disk evolution for the zero-torque cases with $r_{\rm in} \rightarrow 0$ and $r_{\rm in} > 0$ are presented in Sections 3.1 and 3.2, respectively. For the disk with zero torque at the origin, we set a small inner truncation radius of $r_{\rm in} = 10^{-5} r_0$ to evaluate the disk mass, mass accretion rate, and wind mass-loss rate. In contrast, for the disk with a finite inner edge, we adopt $r_{\rm in} = 0.1 r_0$. Our goal is to investigate how this difference in the inner boundary affects the disk evolution and the corresponding physical quantities derived from the solutions.

Figure~\ref{fig:phycomp_caseI} compares the temporal evolution of the disk’s physical quantities for disks extending to the origin and those truncated at a finite inner radius. The mass accretion rate is higher for the disk with a finite inner edge compared to the case with $r_{\rm in} \rightarrow 0$. For the finite inner edge, the peak mass accretion rate increases with $\psi$, whereas it decreases for the disk with $r_{\rm in} \rightarrow 0$. This behavior arises because the mass accretion rate at later times follows $\dot{M} \propto r^{[\sqrt{(1+\psi)^2 + 4\psi/(\lambda - 1)} - (1+\psi)]/4}$, which decreases for smaller inner radii. 
The opposite trends in the accretion rate arise because the radial extent of wind action and the associated mass-loss fraction depend sensitively on $r_{\rm in}$. A smaller $r_{\rm in}$ increases the wind-leveraged region, diverting more mass into outflows, whereas a finite inner edge disk confines the wind’s influence to a narrower zone, where the wind torque instead enhances accretion. Consequently, the wind mass-loss rate, shown in panel (b), is higher for the disk with an inner edge approaching the origin. The ratio of mass accretion to wind-loss rates, presented in panel (c), indicates that the disk loses most of its mass through the wind at higher $\psi$. The evolution of the disk mass becomes nearly identical at higher $\psi$ for both cases—near the origin and at a finite inner radius—whereas a noticeable difference appears when the wind is absent. At high $\psi$, strong winds remove most of the disk mass, reducing the mass inflow toward the inner boundary. This behavior arises because the accretion and wind components are strongly coupled through mass conservation, such that a reduction in the wind mass-loss rate leads to a corresponding increase in the accretion rate.


\subsection{Comparison for three boundary conditions imposed at finite inner edge}

%
\begin{figure}
	\centering
	\subfigure[]{\includegraphics[scale = 0.63]{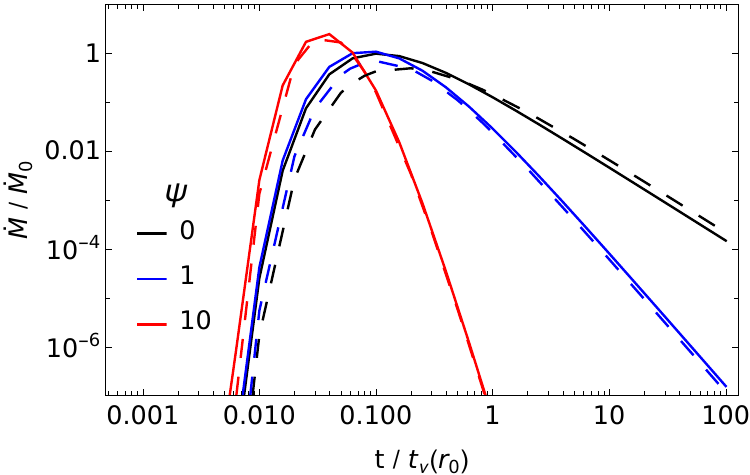}}
	\subfigure[]{\includegraphics[scale = 0.63]{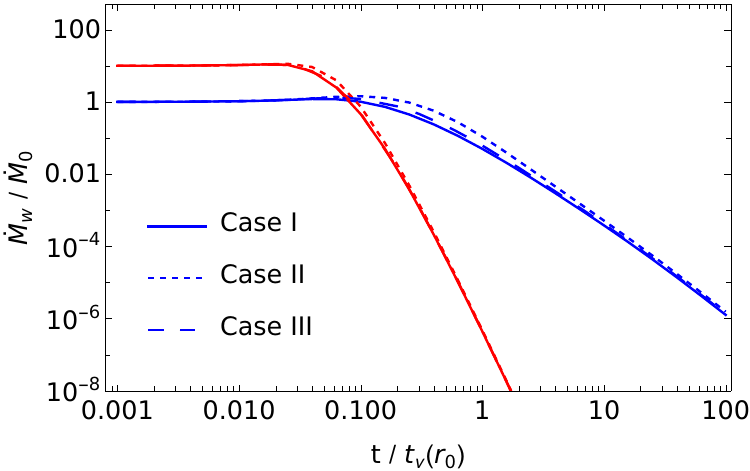}}
	\subfigure[]{\includegraphics[scale = 0.63]{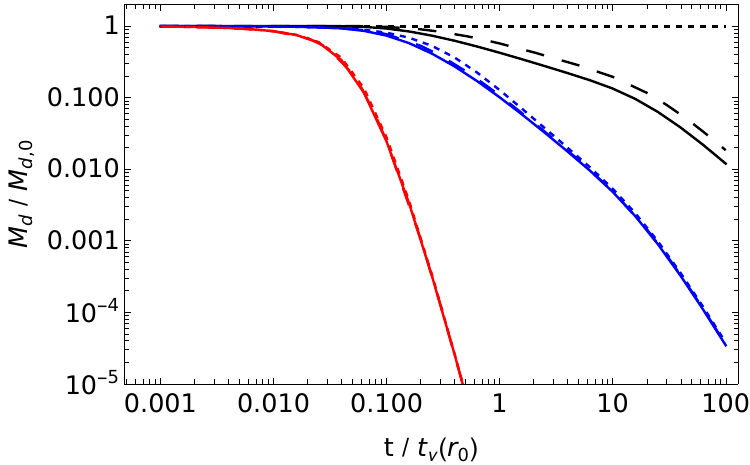}}
	\subfigure[]{\includegraphics[scale = 0.63]{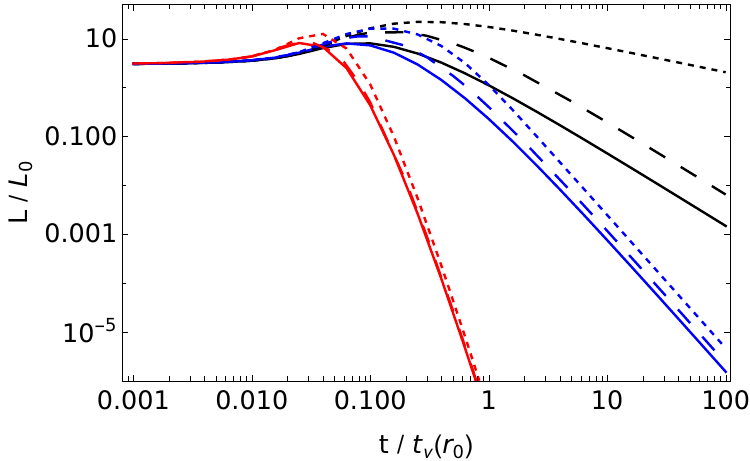}}
	\caption{Comparison of the time evolution of the mass accretion rate (panel a), wind mass loss rate (panel b), disk mass (panel c), and bolometric luminosity (panel d) for the three cases of boundary conditions at a finite inner edge. Case I represents a disk with zero torque at the inner edge, Case II indicates a disk with zero mass accretion rate at the inner edge, and Case III corresponds to a disk with finite torque and finite mass accretion rate at the inner edge. Different colored lines correspond to different values of $\psi$, while different line styles represent the three cases. The short-dashed line is absent in panel (a) due to the zero mass accretion rate boundary condition in Case II. The black line is absent in panel (b) because $\psi = 0$ corresponds to a disk without wind.}
	\label{fig:phycomp} 
\end{figure}

Figure~\ref{fig:phycomp} compares the temporal evolution of the disk quantities for three boundary conditions imposed at $r_{\rm in} = 0.1r_0$. The choice of boundary condition significantly influences the mass accretion rate, mass loss rate, and bolometric luminosity for low values of $\psi$. However, for $\psi \gtrsim 10$, the impact of the boundary condition becomes negligible, particularly at late times.

%


The lower mass accretion rate in Case III (finite torque and finite mass accretion rate at the inner edge) compared to Case I (zero torque) for low values of $\psi$ arises from how the inner boundary condition influences the surface density and angular momentum transport near the inner edge. In Case I, the zero-torque condition enforces a vanishing surface density at the inner boundary, resulting in a steep radial gradient in viscous stress and a relatively high inward mass flux. This boundary condition allows angular momentum to be efficiently transported outward, facilitating accretion. In contrast, in Case III, the surface density at the inner edge is non-zero, and the finite torque alters the angular momentum transport profile near the boundary, thereby reducing the efficiency of inward mass transport and leading to a lower accretion rate. At early times, the higher mass accretion rate in Case I depletes the disk's surface density more rapidly than in Case III. As a result, the mass accretion rate in Case III becomes slightly higher than that of Case I in the absence of wind ($\psi = 0$) at later times. When wind is present, the mass accretion rates in both cases become closely comparable at late times, due to the modestly enhanced mass loss rate in Case III during the intermediate phase of evolution.

The mass-loss rate is highest for the disk with a zero–mass-accretion-rate boundary condition at the inner edge (Case II) and lowest for the disk with a zero-torque boundary condition (Case I). The choice of boundary condition significantly influences the mass-loss rate during intermediate evolutionary stages, particularly for $\psi \leq 1$, whereas the late-time behavior becomes nearly independent of it. The combined effects of mass accretion onto the central object and mass loss through the wind lead to a gradual decrease in the total disk mass. As shown in panel (c), the disk mass remains constant for Case II when $\psi = 0$, because of the zero mass accretion to the central object. However, it decreases with time for $\psi > 0$. At larger values of $\psi$, the temporal evolution of the disk mass becomes insensitive to the boundary condition, as a higher fraction of the disk mass is lost through winds, and both the accretion and wind mass-loss rates follow the same temporal evolution, $\propto t^{-(1+l)}$, across all cases.

The bolometric luminosity is found to be greatest for the disk that imposes a zero–mass-accretion-rate condition at the inner boundary (Case II), whereas it is the lowest for the disk that satisfies a zero-torque condition at the inner boundary (Case I). It is to note that the surface density at the inner edge in Case II is non-zero, indicating the presence of finite torque (see Figure~\ref{fig:zmrinf}). This suggests that disks with finite torque are more luminous than those with zero torque. Case III, which allows both finite torque and mass accretion, exhibits intermediate behavior between Cases I and II. The differences in mass loss rate and bolometric luminosity among the three cases arise from how the inner boundary condition affects energy dissipation in the disk. In Case II, the presence of finite torque allows for non-zero surface density and continued viscous stress and dissipation near the inner edge. Since no mass is accreted onto the central object, all the generated energy is retained in the disk and contributes to radiation and wind, resulting in the highest luminosity and mass loss rate. In contrast, Case I enforces a vanishing surface density at the inner edge and allows free mass inflow, causing some of the disk energy to be lost into the central object and reducing both luminosity and wind strength. Case III, which allows both finite torque and mass accretion, exhibits intermediate behavior between Cases I and II. For $\psi \gtrsim 10$, the wind dominates the evolution, and the influence of the inner boundary condition becomes negligible.

%
\section{Discussion} 
\label{sec:discus}
%

We have constructed Green’s function solutions to the evolution equation for a thin Keplerian accretion disk with a magnetically driven wind, assuming a power-law turbulent viscosity profile of $\nu \propto r^{n}$. These solutions are derived for three types of boundary conditions at the inner radius $r_{\rm in}$—zero viscous torque (case I), zero mass accretion rate (case II), and finite torque with finite mass accretion rate (case III)—considering both $r_{\rm in} = 0$ and finite $r_{\rm in} > 0$. In the absence of the wind where $\psi = 0$, our Green’s function solutions are identical to those derived for a thin disk by Tanaka (2011) \cite{2011MNRAS.410.1007T} for the same boundary conditions. In the presence of the wind, the disk evolution is identical for both zero viscous torque and zero mass accretion rate boundary conditions when $r_{\rm in} = 0$. The boundary condition impacts the disk evolution and the physical quantities such as mass accretion rate, mass loss rate and bolometric luminosity significantly for low $\psi$. However, for $\psi \gtrsim 10$, the impact of the boundary condition becomes negligible, particularly at late times. These solutions provide a means to study how the inner regions of accretion disks evolve over time, particularly in cases where the central object’s finite size has a notable impact on the disk’s observable properties.

Protoplanetary disks are rotationally supported structures of gas and dust surrounding young, typically pre-main-sequence stars, with initial gas masses amounting to a few percent up to about 10\% of the stellar mass \cite{2011ARA&A..49...67W,2022A&A...657A..74F}. Infrared and millimeter surveys reveal that these disks undergo substantial evolution over $\sim$1–10 Myr, consistent with the rapid dispersal of gas and dust during this period \cite{2017RSOS....470114E}. These observations indicate that protoplanetary disks behave as quasi-steady fluid configurations whose long-term evolution is governed by gradual angular momentum redistribution through viscous or magnetically mediated stresses, accompanied by mass loss through disk winds. Consequently, a time-dependent framework that self-consistently incorporates both viscous spreading and magnetically driven outflows—such as the Green’s function formalism developed in this work—is useful for accurately capturing the secular evolution of disk mass and accretion rate over Myr timescales.

The ALMA Survey of Gas Evolution of PROtoplanetary Disks (AGE-PRO) by Zhang et al. (2025)\cite{2025ApJ...989....1Z}, conducted across three nearby star-forming regions—Ophiuchus (0.5–1 Myr), Lupus (1–3 Myr), and Upper Sco (2–6 Myr)—provides key observational constraints on disk evolution. The survey estimates average gas disk masses of $0.0057~M_{\odot}$ for Ophiuchus, $0.000646~M_{\odot}$ for Lupus, and $0.00042~M_{\odot}$ for Upper Sco, with mean gas disk radii ranging from 74 to 110 au. Assuming an initial disk mass of $M_{\rm d,0} = 0.01~M_{\odot}$ and a characteristic radius $r_0 = 1$ au, the surface density normalization constant derived from equation (\ref{eq:mdini}) is $\Sigma_0 = 14222.96~{\rm g~cm^{-2}}$. The surface density evolution shown in panel (a) of Figures \ref{fig:ztrin0}, \ref{fig:zmrinf}, and \ref{fig:finsigmass} illustrates that the disk expands from a compact initial configuration to sizes of several tens to hundreds of astronomical units over a timescale of $t_{\nu}(r_0)$, which can be reasonably approximated as $\sim$1 Myr \cite{2011ARA&A..49..195A}. Panel (c) of Figure~\ref{fig:phycomp} shows that the disk mass decreases over time, with the decay becoming steeper as $\psi$ increases. This indicates that the presence of a magnetically driven wind accelerates disk evolution, thereby shortening the disk’s lifetime.

\begin{figure}
	\centering
	\subfigure[]{\includegraphics[scale = 0.64]{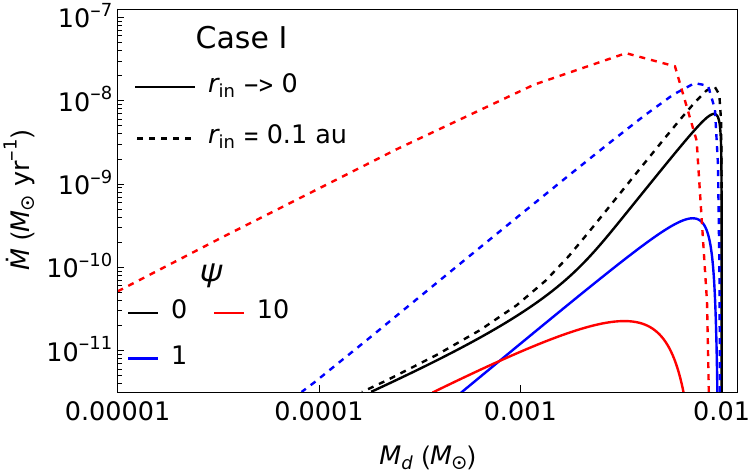}}
	\subfigure[]{\includegraphics[scale = 0.64]{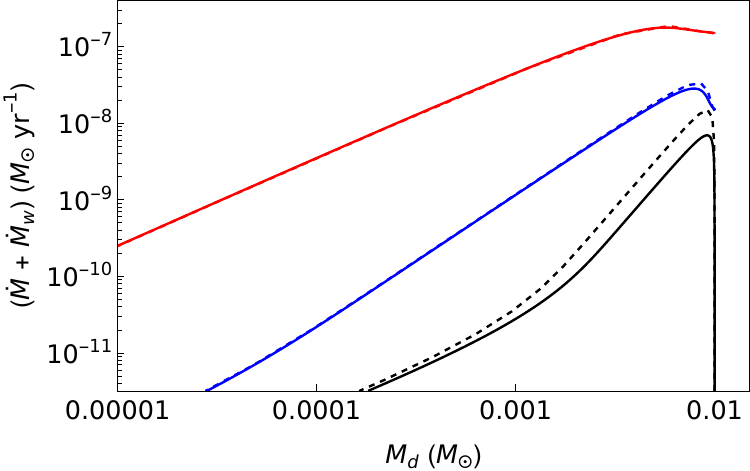}}
	\subfigure[]{\includegraphics[scale = 0.64]{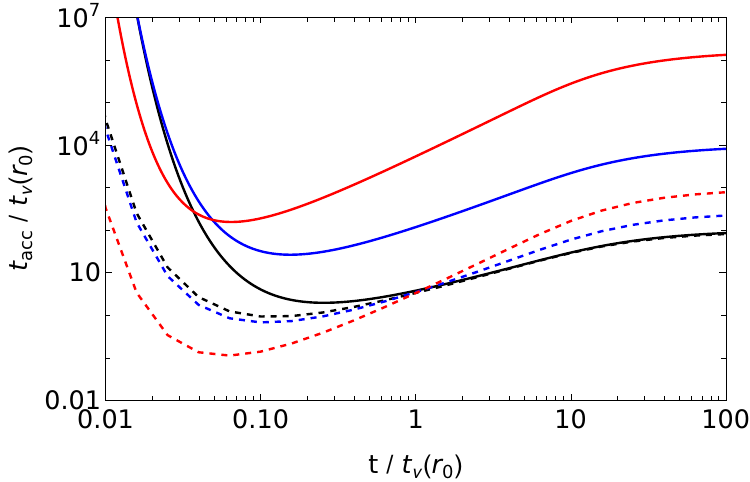}}
	\subfigure[]{\includegraphics[scale = 0.64]{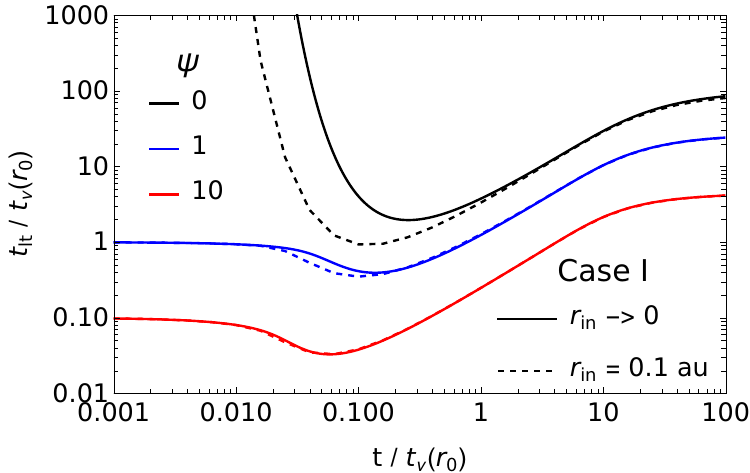}}
	\caption{Evolutionary tracks in the $\dot{M}$–$M_{\rm d}$ and $(\dot{M} + \dot{M}_{\rm w})$–$M_{\rm d}$ planes, along with the accretion timescale ($t_{\rm acc} = M_{\rm d}/\dot{M}$) and the disk lifetime timescale ($t_{\rm lt} = M_{\rm d}/(\dot{M} + \dot{M}_{\rm w})$), are shown for three values of $\psi$: 0 (black), 1 (blue), and 10 (red). These results correspond to Case I, representing the disk with zero torque at the inner boundary. The solid and dashed lines correspond to disks with inner edges at the origin and at a finite radius of 0.1 au, respectively. Panels (a) and (b) show the evolutionary tracks in the $\dot{M}$–$M_{\rm d}$ and $(\dot{M} + \dot{M}_{\rm w})$–$M_{\rm d}$ planes, respectively, while panels (c) and (d) display the accretion and disk lifetime timescales.	}
	\label{fig:MaccMD_caseI} 
\end{figure}
%

\begin{figure}
	\centering
	\subfigure[]{\includegraphics[scale = 0.64]{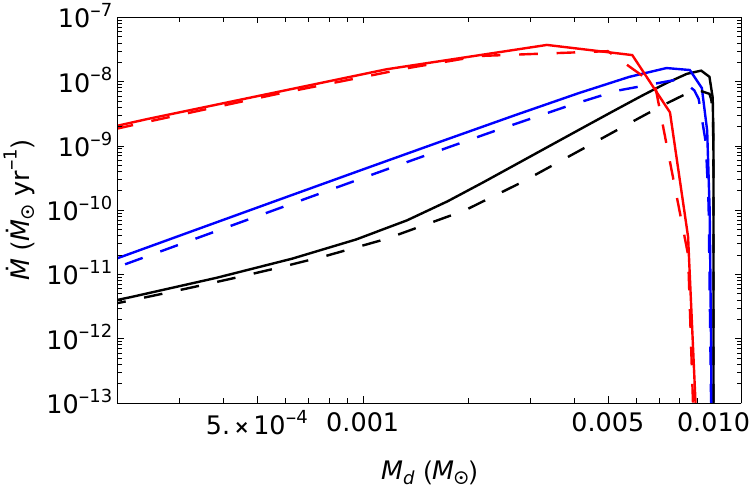}}
	\subfigure[]{\includegraphics[scale = 0.64]{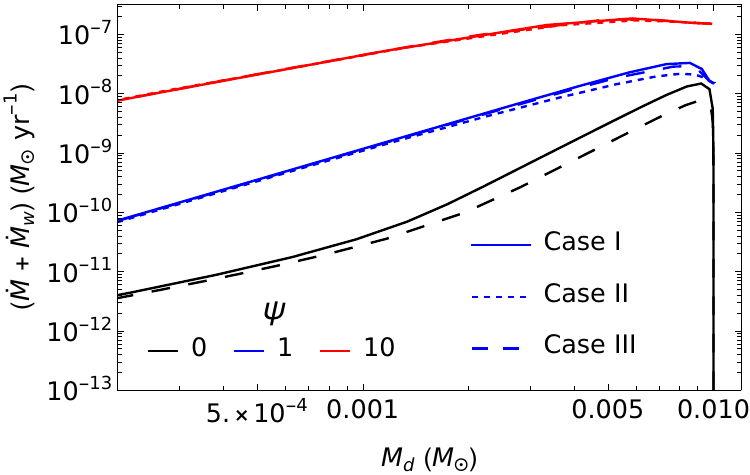}}
	\subfigure[]{\includegraphics[scale = 0.64]{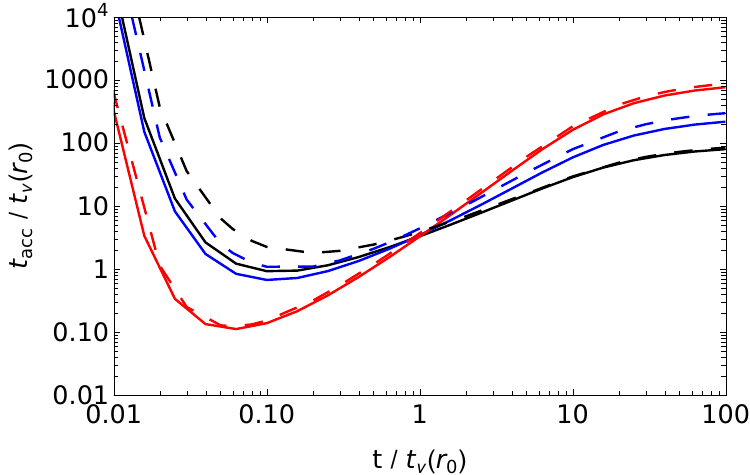}}
	\subfigure[]{\includegraphics[scale = 0.64]{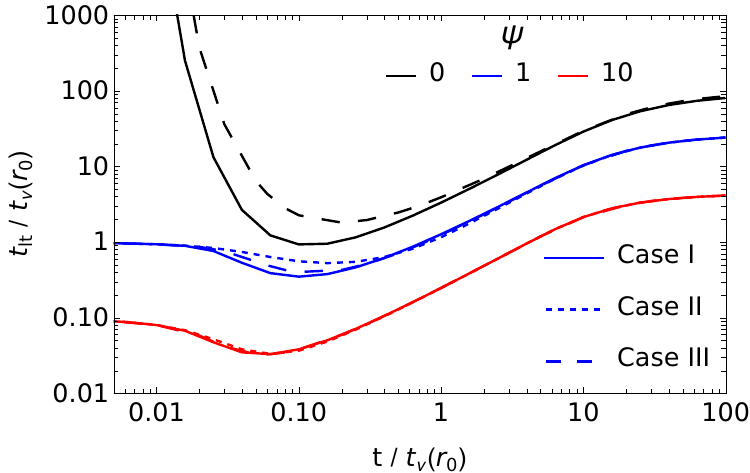}}
	\caption{Evolutionary tracks in the $\dot{M}$–$M_{\rm d}$ and $(\dot{M} + \dot{M}_{\rm w})$–$M_{\rm d}$ planes, along with the accretion timescale ($t_{\rm acc} = M_{\rm d}/\dot{M}$) and the disk lifetime timescale ($t_{\rm lt} = M_{\rm d}/(\dot{M} + \dot{M}_{\rm w})$), are shown for three values of $\psi$: 0 (black), 1 (blue), and 10 (red). The solid, dashed, and long-dashed lines correspond to zero torque (Case I), zero mass accretion rate (Case II), and finite torque with finite mass accretion rate (Case III), respectively, all with the inner edge fixed at 0.1 au. Panels (a) and (b) show the evolutionary tracks in the $\dot{M}$–$M_{\rm d}$ and $(\dot{M} + \dot{M}_{\rm w})$–$M_{\rm d}$ planes, respectively, while panels (c) and (d) present the accretion and disk lifetime timescales.  The short-black dashed line is absent due to the zero mass accretion rate boundary condition in Case II and $\psi = 0$ corresponding to no wind.	}
	\label{fig:MaccMD_caseall} 
\end{figure}

The evolutionary tracks in the $\dot{M}$–$M_{\rm d}$ plane serve as key diagnostics of disk evolution, linking mass accretion to the depletion of disk material. Observationally, $\dot{M}$ is inferred from the accretion luminosity, estimated through ultraviolet or optical continuum excess or from emission-line luminosities such as H$\alpha$ or Ca II \cite{1998ApJ...509..802C,2014A&A...561A...2A}. The disk mass $M_{\rm d}$ is derived from (sub-)millimeter dust continuum observations, assuming a typical gas-to-dust ratio of 100:1 \cite{2013ApJ...771..129A,2016ApJ...828...46A}. Together, these measurements provide empirical $\dot{M}$–$M_{\rm d}$ relations that constrain models of angular momentum transport and mass loss in protoplanetary disks. Panel (a) of Figures~\ref{fig:MaccMD_caseI} and~\ref{fig:MaccMD_caseall} show the $\dot{M}$–$M_{\rm d}$ evolutionary tracks for Case I (with zero and finite inner edges) and for all boundary conditions with finite inner edges, respectively. The corresponding temporal evolution of the mass accretion rate and disk mass are presented in Figures~\ref{fig:phycomp_caseI} and~\ref{fig:phycomp}. For disks with finite $r_{\rm in}$, the evolutionary tracks differ between boundary conditions at low $\psi$, while this difference becomes negligible at higher $\psi$. At later times (i.e., lower $M_{\rm d}$), $\dot{M}$ exhibits a shallower dependence on $M_{\rm d}$ as $\psi$ increases, consistent with the results of Tabone et al. (2022) \cite{2022MNRAS.512.2290T}. Since the disk also loses mass through winds, the total depletion rate is given by $\dot{M} + \dot{M}_{\rm w}$, whose evolution is shown in panel (b) of Figures~\ref{fig:MaccMD_caseI} and~\ref{fig:MaccMD_caseall}. The total depletion rate exceeds the accretion rate alone by nearly an order of magnitude at later times. For Case I, the total depletion rate evolves similarly to the disk mass for both zero and finite inner edges as $\psi$ increases, with differences apparent only at early times (when the disk mass is higher) for low $\psi$. Similarly, for disks with a finite inner edge, the evolutionary tracks differ among the three boundary conditions—zero torque, zero accretion rate, and finite torque with finite accretion rate—most prominently near the peak accretion phase for $\psi \lesssim 1$, but these differences diminish as $\psi$ increases, indicating that wind-driven angular momentum transport dominates over the influence of the inner boundary condition in the strong-wind regime.

%

The accretion timescale, defined as $t_{\rm acc} = M_{\rm d}/\dot{M}$, represents the time required for the disk to deplete its mass solely through accretion onto the central star. In contrast, the disk lifetime, $t_{\rm lt} = M_{\rm d}/(\dot{M} + \dot{M}_{\rm w})$, accounts for both accretion and wind-driven mass loss, providing a more realistic measure of the total depletion timescale. Observations suggest that typical protoplanetary disks persist for a few million years, consistent with the gradual decline of accretion rate and disk mass over time \cite{2010A&A...510A..72F,2011ARA&A..49..195A,2017RSOS....470114E}. Panels (c) and (d) of Figures~\ref{fig:MaccMD_caseI} and~\ref{fig:MaccMD_caseall} shows the accretion and disk lifetimes in units of $t_{\nu}(r_0)$ for various cases. The accretion timescale $t_{\rm acc}$ is shorter at early times and becomes longer at later times as $\psi$ increases. This trend arises because stronger winds (larger $\psi$) enhance angular momentum removal, leading to higher accretion rates and faster disk depletion initially, followed by lower accretion rates at later stages (see Figure~\ref{fig:phycomp}). For a representative viscous timescale of $t_{\nu}(r_0) = 1$ Myr, $t_{\rm acc}$ ranges from $\sim$100 Myr for $\psi = 0$ to $\sim10^9$ yr for $\psi = 10$. The disk lifetime $t_{\rm lt}$, however, becomes substantially shorter in the presence of winds, decreasing with increasing $\psi$ due to enhanced mass loss by the wind. For $t_{\nu}(r_0) = 1$ Myr, $t_{\rm lt}$ varies from a few Myr for $\psi = 10$ to about 100 Myr for $\psi = 0$. The impact of the inner boundary condition is noticeable only at early times for $\psi \leq 1$, while it becomes negligible for higher $\psi$, indicating that strong winds dominate the long-term disk evolution.

The Green’s function solution adopted in this work provides a robust framework to trace the temporal evolution of disk properties for arbitrary initial mass distributions, enabling a direct comparison between cases with and without wind-driven angular momentum loss. For weak winds ($\psi \leq 1$), the evolution is sensitive to the inner boundary condition and shows noticeable differences near the peak accretion phase, while for stronger winds, the evolutionary behavior converges, indicating that mass and angular momentum removal through winds dominates the disk’s evolution. The decreasing disk lifetime with increasing $\psi$ implies that strong winds can rapidly deplete disk material, influencing the timescales for planet formation and migration. These theoretical results are consistent with the observed decline of accretion rates and disk fractions over a few million years in young stellar clusters \cite{2010A&A...510A..72F,2011ARA&A..49..195A,2017RSOS....470114E}, suggesting that the wind-driven evolution described by the Green’s function solution offers a plausible framework to investigate the observed diversity in disk dispersal timescales.

As seen in our results (panel (a) in Figures \ref{fig:ztrin0}, \ref{fig:zmrinf} and \ref{fig:finsigmass}), increasing $\psi$ removes more angular momentum directly from the disk, which limits the outward radial flow of gas and keeps the disk more compact compared to no-wind case. This framework can therefore be used to study how winds influence the evolution of disk size over time, and to compare theoretical predictions with observed disk radii in protoplanetary systems \cite{2013ApJ...771..129A,2016ApJ...828...46A,2025ApJ...989....1Z}. By applying this approach to disks with varying initial masses, inner radii, and wind parameters, the Green’s function solution can explore how winds influence the growth, truncation, and overall spatial extent of disks over their evolutionary timescales. A more detailed analysis along these lines, including comparisons with observed disk radius–age distributions, will be pursued in future work. 

Apart from protoplanetary disks, the derived Green’s function solution is also applicable to the long-term evolution of accretion disks in a variety of astrophysical systems, such as those around stars, X-ray binaries (XRBs), and active galactic nuclei (AGNs). In particular, the Green’s function with a zero mass accretion rate at a finite inner radius (Case II) is relevant for disks truncated by a strong stellar magnetosphere, such as those around magnetized neutron stars. When the magnetospheric radius exceeds the corotation radius, centrifugal forces inhibit accretion onto the central object, leading to the propeller regime \cite{1993ApJ...402..593S,1995MNRAS.275..244L,2023MNRAS.524.1727C}. In this regime, the accretion onto the star effectively ceases, but the disk continues to evolve through viscous processes, redistributing its mass and angular momentum outward. Thus, the Green’s function model provides a useful theoretical framework for investigating the secular viscous evolution of magnetically truncated disks in both stellar and compact object systems.

MHD winds originating from the accretion disk give rise to an extended, ionized atmosphere that interacts with the disk’s radiation field. This interaction photoionizes the wind material, with the resulting ionization structure determined by the wind’s density, velocity, and temperature—parameters governed by the underlying accretion-ejection process. Blueshifted absorption lines of highly ionized species such as Fe XXV and Fe XXVI serve as key diagnostics of wind velocity, ionization state, and column density, and have been widely observed in AGNs \cite{2010A&A...521A..57T,2016AN....337..410T,2023A&A...679A..73G} and XRBs \cite{2012MNRAS.422L..11P,2013AdSpR..52..732N,2024A&A...687A...2D}. These lines arise from photoionized gas irradiated by the disk’s X-ray emission and are sensitive to changes in accretion and outflow rates. Moreover, re-emission from the photoionized wind can produce observable optical and UV emission lines, provided the wind's density and temperature are sufficiently high, and  the optical depth permits efficient reprocessing of the incident radiation without trapping the emitted photons \cite{1997ApJ...474...91M,2013A&A...551L...6K,2020MNRAS.494.4914P}. The Green’s function solutions developed in this work offer a time-dependent framework for modeling accretion disks with MHD winds for any initial surface density distribution under different boundary conditions, enabling studies of the evolving disk luminosity and wind structure. This can be applied in future work to compute time-varying photoionization and emission spectra from accretion-driven systems.

%
\section{Summary and conclusion}
\label{sec:sumcon}
%

We derived Green’s function solutions describing the temporal evolution of a thin Keplerian accretion disk subject to a magnetically driven wind, assuming a power-law viscosity profile $\nu \propto r^{n}$. The solutions were obtained for three types of inner boundary conditions—zero torque (Case I), zero accretion rate (Case II), and finite torque with finite accretion (Case III)—for both $r_{\rm in}=0$ and finite $r_{\rm in}>0$. The disk evolution in this framework is governed by three key parameters: $n$, which characterizes the viscosity profile; $\psi$, which quantifies the vertical magnetic stress; and $\lambda$, which measures the efficiency of angular momentum extraction by the wind. Assuming an initial Dirac delta surface density distribution, we computed the time-dependent evolution of the disk surface density. Our main conclusions are summarized as follows:

\begin{enumerate}

\item Our derived Green’s function solution reduces to that obtained by Tanaka (2011)\cite{2011MNRAS.410.1007T} in the absence of wind, demonstrating the consistency of our formulation with previous analytical results.

\item In the zero-torque case (Case I), the mass accretion rate in the absence of wind follows the well-known relation $\dot{M} \propto t^{-3/2}$ at late times for $n=1$, consistent with the classical viscous evolution solution of Lynden-Bell \& Pringle (1974)\cite{1974MNRAS.168..603L}. When a magnetically driven wind is included, the accretion rate declines more steeply with time, following $\dot{M} \propto t^{-(1+l)}$, where $l = \sqrt{(1+\psi)^2 + 4\psi/(\lambda - 1)} / (4 - 2n)$. The mass-loss rate through the wind, $\dot{M}_{\rm w}$, exhibits a similar temporal dependence at late times, though it differs from the accretion rate evolution at earlier stages.

\item For the boundary conditions corresponding to zero mass accretion rate (Case II) and finite torque with finite accretion (Case III), both the mass accretion rate and the wind mass-loss rate exhibit the same late-time temporal evolution as in Case I when the wind is present, following $\dot{M}, \dot{M}_{\rm w} \propto t^{-(1+l)}$.

\item The choice of boundary condition has a substantial impact on the mass accretion rate, wind mass-loss rate, and bolometric luminosity when $\psi \lesssim 1$. However, this influence diminishes for larger values of $\psi$, where the disk evolution becomes increasingly insensitive to the boundary condition (see Figure \ref{fig:phycomp}).

\item The bolometric luminosity is highest for the disk with a zero–mass-accretion-rate inner boundary (Case II) and lowest for the zero-torque case (Case I), indicating that disks with finite inner torque are more luminous. Case III, which permits both finite torque and accretion, shows intermediate luminosity between Cases I and II.

\item We applied our Green’s function solution to examine the protoplanetary disks evolution in the $\dot{M}$–$M_{\rm d}$ and $(\dot{M} + \dot{M}_{\rm w})$–$M_{\rm d}$ planes (see Figures~\ref{fig:MaccMD_caseI} and \ref{fig:MaccMD_caseall}). The mass accretion rate exhibits a positive correlation with disk mass at later stages (i.e., lower $M_{\rm d}$), with the $\dot{M}$–$M_{\rm d}$ relation becoming progressively shallower as $\psi$ increases. The total mass ejection rate, $(\dot{M} + \dot{M}_{\rm w})$, follows a similar trend since both the accretion and wind mass-loss rates exhibit similar temporal dependence.

The disk accretion timescale, $t_{\rm acc} = M_{\rm d}/\dot{M}$, increases with $\psi$ at later times due to the suppression of mass accretion by stronger winds, and thus does not accurately reflect the true disk lifetime when winds are present. Instead, the disk lifetime, defined as $t_{\rm lt} = M_{\rm d}/(\dot{M} + \dot{M}_{\rm w})$ (see Figures~\ref{fig:MaccMD_caseI} and \ref{fig:MaccMD_caseall}), decreases with increasing $\psi$, because the total mass ejection rate exceeds the accretion rate. The boundary condition affects both the evolutionary tracks and characteristic timescales at low $\psi$, but its influence becomes negligible at higher $\psi$, indicating that strong magnetically driven winds dominate the long-term disk evolution and substantially reduce the mass flow toward the inner boundary.

\end{enumerate}

\section*{Acknowledgment}

We thank the referee for constructive suggestions that have improved the paper. We are grateful to Kimitake Hayasaki and Takeru K. Suzuki for their insightful discussions and valuable suggestions, which greatly benefited this study.


\appendix

\section{Integral transforms}
\label{sec:inttrans}

The Hankel transform pair of order $l$ satisfies 
\begin{eqnarray}
	\phi_{l}(x) = \int_0^{\infty} \Phi_{l}(k) J_{l}(k x) k \, \diff k, \label{eq:Hankel}\\
	\Phi_{l}(k) = \int_0^{\infty} \phi_{l}(x) J_{l}(k x) x \, \diff x,\label{eq:Hankel1}
\end{eqnarray}
where $\phi$ and $\Phi$ are arbitrary functions. 

The Weber integral transform pair is given by \cite{Titchmarsh1924}
\begin{eqnarray}\label{eq:webint}
	\phi_{l}(x) &&=\int_{0}^{\infty} \Phi_{l}(k_1) \left[\frac{J_{l}(k_1 x)Y_{l}(k_1) - Y_{l}(k_1 x)J_{l}(k_1)}{J_{l}^2(k_1)+Y_{l}^2(k_1)}\right] k_1 \, \diff k_1, \\
	\Phi_{l}(k_1) &&= \int_{1}^{\infty} \phi_{l}(x) \left[J_{l}(k_1 x)Y_{l}(k_1) - Y_{l}(k_1 x)J_{l}(k_1)\right] x \, \diff x.\label{eq:webint1}
\end{eqnarray} 

A generalized Weber transform is \cite{ZHANG2007116},
\begin{eqnarray}
	\phi_{l}(x) =&& \int_{0}^{\infty} \frac{W_{l}(k_1,x,a,b)}{Q_{l}^2(k_1,a,b)} \Phi_{l}(k_1) k_1 \, \diff k_1, \label{eq:gweber}\\
	\Phi_{l}(k_1) =&& \int_{1}^{\infty} W_{l}(k_1,x,a,b) \phi_{l}(x) x \, \diff x, \label{eq:gweber1}
\end{eqnarray}
where 
\begin{eqnarray}
	W_{l}(k_1,x,a,b) &=& J_{l}(k_1 x)\left[(a-lb)Y_{l}(k_1)+ b k_1 Y_{l-1}(k_1) \right] - Y_{l}(k_1 x)\left[(a-lb)J_{l}(k_1)+ b k_1 J_{l-1}(k_1) \right] \nonumber \\ \label{eq:Wlfn}\\ 
	Q_{l}^2(k_1,a,b) &=& \left[(a-lb)Y_{l}(k_1)+ b k_1 Y_{l-1}(k_1) \right]^2+\left[(a-lb)J_{l}(k_1)+ b k_1 J_{l-1}(k_1) \right]^2. \label{eq:Qlfn}
\end{eqnarray}

\section{Comparison of Asymptotic Disk Surface Density with Zero-Torque Boundary to Tabone et al. (2022)\cite{2022MNRAS.512.2290T}}
\label{sec:appcomp}

Tabone et al. (2022) \cite{2022MNRAS.512.2290T} adopted $\alpha_{\rm SS} c_s^2 \propto r^{-3/2+\gamma}$ and $\alpha_{\rm DW} c_s^2 \propto r^{-3/2+\gamma}$ in their model, where $\alpha_{\rm SS}$ and $\alpha_{\rm DW}$ are the radial turbulent and vertical disk-wind stress $\alpha-$parameters. It is to note that $\gamma = n$ as shown above equation (\ref{eq:sigevn2}). Their self-similar solution (Appendix C in Tabone et al. 2022) reads:
\begin{equation}
\Sigma(r,t) = A(0) \left(1+\frac{t}{t_{\nu,0}}\right)^{-(1+l)} \left(\frac{r}{r_{c}(0)}\right)^{-\gamma+(1-\gamma/2)l-(1+\psi)/4} \exp\left(-\left\{1+\frac{t}{t_{\nu,0}}\right\}^{-1}\left(\frac{r}{r_{c}(0)}\right)^{2-\gamma}\right),
\end{equation} 
where $A(0)$ is a constant, $r_{c}(0)$ is the characteristic radius at $t=0$, $l$ is given by equation (\ref{eq:l}), and $t_{\nu,0}$ is the viscous timescale at $r_{c}(0)$ (see equation C3 in \cite{2022MNRAS.512.2290T}). For $t \gg t_{\nu,0}$, the asymptotic solution becomes
\begin{equation}\label{eq:taboneasmp}
\Sigma(r,t) = A(0) \left(\frac{t}{t_{\nu,0}}\right)^{-(1+l)} \left(\frac{r}{r_{c}(0)}\right)^{-n+\beta l-(1+\psi)/4} \exp\left(-\left\{\frac{t}{t_{\nu,0}}\right\}^{-1}\left(\frac{r}{r_{c}(0)}\right)^{2\beta}\right),
\end{equation} 
where we have used $\gamma = n$ and equation (\ref{eq:delta}).

The surface density for a disk with zero torque at $r_{\rm in} = 0$ is given by equation (\ref{eq:sigzt0}): 
\begin{multline}
	\Sigma(r,t)=\frac{\Sigma_0}{6 \beta}\left(\frac{t}{t_{\nu}(r_0)}\right)^{-1}\left(\frac{r}{r_0}\right)^{-n-(1+\psi)/4} I_{l}\left[\frac{1}{6 \beta^2}\left(\frac{t}{t_{\nu}(r_0)}\right)^{-1}\left(\frac{r}{r_0}\right)^{\beta}\right]\\ \exp\left[-\frac{1}{12\beta^2}\left(\frac{t}{t_{\nu}(r_0)}\right)^{-1}\left\{1+\left(\frac{r}{ r_0}\right)^{2\beta}\right\}\right].
\end{multline}
For $t \gg t_{\nu}(r_0)$, using $I_l(z) \sim (z/2)^l / \Gamma(1+l)$ for $z \ll 1$, the asymptotic solution becomes
\begin{multline}\label{eq:oursxt}
\Sigma(r,t)=\frac{\Sigma_0}{6 \beta(12\beta^2)^l \Gamma(1+l)}\left(\frac{t}{t_{\nu}(r_0)}\right)^{-(1+l)}\left(\frac{r}{r_0}\right)^{-n+\beta l-(1+\psi)/4} \\ \exp\left[-\frac{1}{12\beta^2}\left(\frac{t}{t_{\nu}(r_0)}\right)^{-1}\left\{1+\left(\frac{r}{ r_0}\right)^{2\beta}\right\}\right].
\end{multline}
Here, the exponential term represents the radial cutoff of the disk: for $r < r_0$, $\left(r/r_0\right)^{2\beta}$ is small and the exponential is nearly unity, so the surface density is dominated by the power-law part. For $r > r_0$, $\left(r/r_0\right)^{2\beta}$ grows and the exponential decays rapidly, giving the outer disk a sharp cutoff. Thus, keeping the dominant contribution within the exponential captures the radial structure of the disk, such that equation~(\ref{eq:oursxt}) reduces to
\begin{multline}\label{eq:ourasmp}
\Sigma(r,t)\simeq\frac{\Sigma_0}{6 \beta(12\beta^2)^l \Gamma(1+l)}\left(\frac{t}{t_{\nu}(r_0)}\right)^{-(1+l)}\left(\frac{r}{r_0}\right)^{-n+\beta l-(1+\psi)/4} \\ \exp\left[-\frac{1}{12\beta^2}\left(\frac{t}{t_{\nu}(r_0)}\right)^{-1}\left(\frac{r}{ r_0}\right)^{2\beta}\right].
\end{multline}
Our asymptotic solution in equation (\ref{eq:ourasmp}) matches that of Tabone et al. (2022) (equation \ref{eq:taboneasmp}) under the identifications $r_0 = r_c(0)$, $t_{\nu}(r_0) = 12 \beta^2 t_{\nu,0}$, and $\Sigma_0 = A(0) \Gamma(1+l)/(2 \beta)$.


%

\vspace{0.2cm}
\noindent

\let\doi\relax
\bibliographystyle{ptephy}
\bibliography{reference}

\begin{thebibliography}{10}

\bibitem{2002apa..book.....F}
Juhan {Frank}, Andrew {King}, and Derek~J. {Raine},
\newblock {\em {Accretion Power in Astrophysics: Third Edition}},
\newblock  (Cambridge University Press, 2002).

\bibitem{2008bhad.book.....K}
S.~{Kato}, J.~{Fukue}, and S.~{Mineshige},
\newblock {\em {Black-Hole Accretion Disks --- Towards a New Paradigm ---}},
\newblock  (Kyoto University Press (Kyoto, Japan), 2008).

\bibitem{1973A&A....24..337S}
N.~I. {Shakura} and R.~A. {Sunyaev}, \aap, {\bf 24}, 337--355 (January 1973).

\bibitem{1974MNRAS.168..603L}
D.~{Lynden-Bell} and J.~E. {Pringle}, \mnras, {\bf 168}, 603--637 (September
  1974).

\bibitem{2011MNRAS.410.1007T}
Takamitsu {Tanaka}, \mnras, {\bf 410}(2), 1007--1017 (January 2011),
  {{arXiv:1007.4474}}.

\bibitem{Velikhov1959}
E.~P. {Velikhov}, Zh. Eksp. Teor. Fiz., {\bf 36}, 1398 (1959).

\bibitem{1991ApJ...376..214B}
Steven~A. {Balbus} and John~F. {Hawley}, \apj, {\bf 376}, 214 (July 1991).

\bibitem{1998RvMP...70....1B}
Steven~A. {Balbus} and John~F. {Hawley}, Reviews of Modern Physics, {\bf
  70}(1), 1--53 (January 1998).

\bibitem{1999ApJ...521..650B}
Steven~A. {Balbus} and John C.~B. {Papaloizou}, \apj, {\bf 521}(2), 650--658
  (August 1999),  {{arXiv:astro-ph/9903035}}.

\bibitem{2009ApJ...691L..49S}
Takeru~K. {Suzuki} and Shu-ichiro {Inutsuka}, \apjl, {\bf 691}(1), L49--L54
  (January 2009),  {{arXiv:0812.0844}}.

\bibitem{2014ApJ...784..121S}
Takeru~K. {Suzuki} and Shu-ichiro {Inutsuka}, \apj, {\bf 784}(2), 121 (April
  2014),  {{arXiv:1309.6916}}.

\bibitem{2013ApJ...765..149C}
Xinwu {Cao} and Hendrik~C. {Spruit}, \apj, {\bf 765}(2), 149 (March 2013),
  {{arXiv:1301.4543}}.

\bibitem{2019ApJ...872..149L}
Jiawen {Li} and Xinwu {Cao}, \apj, {\bf 872}(2), 149 (February 2019),
  {{arXiv:1901.10103}}.

\bibitem{1982MNRAS.199..883B}
R.~D. {Blandford} and D.~G. {Payne}, \mnras, {\bf 199}, 883--903 (June 1982).

\bibitem{2016A&A...596A..74S}
Takeru~K. {Suzuki}, Masahiro {Ogihara}, Alessandro {Morbidelli}, Aur{\'e}lien
  {Crida}, and Tristan {Guillot}, \aap, {\bf 596}, A74 (December 2016),
  {{arXiv:1609.00437}}.

\bibitem{2024ApJ...975...94T}
Mageshwaran {Tamilan}, Kimitake {Hayasaki}, and Takeru~K. {Suzuki}, \apj, {\bf
  975}(1), 94 (November 2024),  {{arXiv:2312.15415}}.

\bibitem{2022MNRAS.512.2290T}
Beno{\^\i}t {Tabone}, Giovanni~P. {Rosotti}, Alexander~J. {Cridland}, Philip~J.
  {Armitage}, and Giuseppe {Lodato}, \mnras, {\bf 512}(2), 2290--2309 (May
  2022),  {{arXiv:2111.10145}}.

\bibitem{2024MNRAS.528.3294S}
Mohsen {Shadmehri} and Fazeleh {Khajenabi}, \mnras, {\bf 528}(2), 3294--3303
  (February 2024),  {{arXiv:2401.09565}}.

\bibitem{2011ARA&A..49...67W}
Jonathan~P. {Williams} and Lucas~A. {Cieza}, \araa, {\bf 49}(1), 67--117
  (September 2011),  {{arXiv:1103.0556}}.

\bibitem{2022A&A...657A..74F}
Riccardo {Franceschi}, Tilman {Birnstiel}, Thomas {Henning}, Paola {Pinilla},
  Dmitry {Semenov}, and Apostolos {Zormpas}, \aap, {\bf 657}, A74 (January
  2022),  {{arXiv:2110.09406}}.

\bibitem{2010A&A...510A..72F}
D.~{Fedele}, M.~E. {van den Ancker}, Th. {Henning}, R.~{Jayawardhana}, and
  J.~M. {Oliveira}, \aap, {\bf 510}, A72 (February 2010),  {{arXiv:0911.3320}}.

\bibitem{2025PTEP.2025h3E01T}
Mageshwaran {Tamilan}, Kimitake {Hayasaki}, and Takeru~K. {Suzuki}, Progress of
  Theoretical and Experimental Physics, {\bf 2025}(8), 083E01 (August 2025),
  {{arXiv:2502.12549}}.

\bibitem{2024arXiv241100298T}
Mageshwaran {Tamilan}, Kimitake {Hayasaki}, and Takeru~K. {Suzuki}, arXiv
  e-prints, page arXiv:2411.00298 (October 2024),  {{arXiv:2411.00298}}.

\bibitem{2017RSOS....470114E}
Barbara {Ercolano} and Ilaria {Pascucci}, Royal Society Open Science, {\bf
  4}(4), 170114 (April 2017),  {{arXiv:1704.00214}}.

\bibitem{2025ApJ...989....1Z}
Ke~{Zhang}, Laura~M. {P{\'e}rez}, Ilaria {Pascucci}, Paola {Pinilla}, Lucas~A.
  {Cieza}, John {Carpenter}, Leon {Trapman}, Dingshan {Deng}, Carolina
  {Agurto-Gangas}, Anibal {Sierra}, Nicol{\'a}s~T. {Kurtovic}, Dary~A.
  {Ruiz-Rodriguez}, Miguel {Vioque}, James {Miley}, Beno{\^\i}t {Tabone},
  Camilo {Gonz{\'a}lez-Ruilova}, Rossella {Anania}, Giovanni~P. {Rosotti},
  Estephani {TorresVillanueva}, Michiel~R. {Hogerheijde}, Kamber {Schwarz}, and
  Aleksandra {Kuznetsova}, \apj, {\bf 989}(1), 1 (August 2025),
  {{arXiv:2506.10719}}.

\bibitem{2011ARA&A..49..195A}
Philip~J. {Armitage}, \araa, {\bf 49}(1), 195--236 (September 2011),
  {{arXiv:1011.1496}}.

\bibitem{1998ApJ...509..802C}
Nuria {Calvet} and Erik {Gullbring}, \apj, {\bf 509}(2), 802--818 (December
  1998).

\bibitem{2014A&A...561A...2A}
J.~M. {Alcal{\'a}}, A.~{Natta}, C.~F. {Manara}, L.~{Spezzi}, B.~{Stelzer},
  A.~{Frasca}, K.~{Biazzo}, E.~{Covino}, S.~{Randich}, E.~{Rigliaco},
  L.~{Testi}, F.~{Comer{\'o}n}, G.~{Cupani}, and V.~{D'Elia}, \aap, {\bf 561},
  A2 (January 2014),  {{arXiv:1310.2069}}.

\bibitem{2013ApJ...771..129A}
Sean~M. {Andrews}, Katherine~A. {Rosenfeld}, Adam~L. {Kraus}, and David~J.
  {Wilner}, \apj, {\bf 771}(2), 129 (July 2013),  {{arXiv:1305.5262}}.

\bibitem{2016ApJ...828...46A}
M.~{Ansdell}, J.~P. {Williams}, N.~{van der Marel}, J.~M. {Carpenter},
  G.~{Guidi}, M.~{Hogerheijde}, G.~S. {Mathews}, C.~F. {Manara}, A.~{Miotello},
  A.~{Natta}, I.~{Oliveira}, M.~{Tazzari}, L.~{Testi}, E.~F. {van Dishoeck},
  and S.~E. {van Terwisga}, \apj, {\bf 828}(1), 46 (September 2016),
  {{arXiv:1604.05719}}.

\bibitem{1993ApJ...402..593S}
H.~C. {Spruit} and Ronald~E. {Taam}, \apj, {\bf 402}, 593 (January 1993).

\bibitem{1995MNRAS.275..244L}
R.~V.~E. {Lovelace}, M.~M. {Romanova}, and G.~S. {Bisnovatyi-Kogan}, \mnras,
  {\bf 275}(2), 244--254 (July 1995),  {{arXiv:astro-ph/9412030}}.

\bibitem{2023MNRAS.524.1727C}
Sercan {{\c{C}}{\i}k{\i}nto{\u{g}}lu} and K.~Yavuz {Ek{\c{s}}i}, \mnras, {\bf
  524}(2), 1727--1734 (September 2023),  {{arXiv:2211.12945}}.

\bibitem{2010A&A...521A..57T}
F.~{Tombesi}, M.~{Cappi}, J.~N. {Reeves}, G.~G.~C. {Palumbo}, T.~{Yaqoob},
  V.~{Braito}, and M.~{Dadina}, \aap, {\bf 521}, A57 (October 2010),
  {{arXiv:1006.2858}}.

\bibitem{2016AN....337..410T}
F.~{Tombesi}, Astronomische Nachrichten, {\bf 337}(4-5), 410 (May 2016),
  {{arXiv:1603.01235}}.

\bibitem{2023A&A...679A..73G}
M.~{Giustini}, P.~{Rodr{\'\i}guez Hidalgo}, J.~N. {Reeves}, G.~{Matzeu},
  V.~{Braito}, M.~{Eracleous}, G.~{Chartas}, N.~{Schartel}, C.~{Vignali}, P.~B.
  {Hall}, T.~{Waters}, G.~{Ponti}, D.~{Proga}, M.~{Dadina}, M.~{Cappi},
  G.~{Miniutti}, and L.~{de Vries}, \aap, {\bf 679}, A73 (November 2023),
  {{arXiv:2306.05469}}.

\bibitem{2012MNRAS.422L..11P}
G.~{Ponti}, R.~P. {Fender}, M.~C. {Begelman}, R.~J.~H. {Dunn}, J.~{Neilsen},
  and M.~{Coriat}, \mnras, {\bf 422}(1), L11--L15 (May 2012),
  {{arXiv:1201.4172}}.

\bibitem{2013AdSpR..52..732N}
Joey {Neilsen}, Advances in Space Research, {\bf 52}(4), 732--739 (August
  2013),  {{arXiv:1304.6091}}.

\bibitem{2024A&A...687A...2D}
Sudeb~Ranjan {Datta}, Susmita {Chakravorty}, Jonathan {Ferreira},
  Pierre-Olivier {Petrucci}, Timothy~R. {Kallman}, Jonatan {Jacquemin-Ide},
  Nathan {Zimniak}, Joern {Wilms}, Stefano {Bianchi}, Maxime {Parra}, and
  Ma{\"\i}ca {Clavel}, \aap, {\bf 687}, A2 (July 2024),  {{arXiv:2403.13077}}.

\bibitem{1997ApJ...474...91M}
N.~{Murray} and J.~{Chiang}, \apj, {\bf 474}(1), 91--103 (January 1997).

\bibitem{2013A&A...551L...6K}
W.~{Kollatschny} and M.~{Zetzl}, \aap, {\bf 551}, L6 (March 2013),
  {{arXiv:1301.7704}}.

\bibitem{2020MNRAS.494.4914P}
Edward~J. {Parkinson}, Christian {Knigge}, Knox~S. {Long}, James~H. {Matthews},
  Nick {Higginbottom}, Stuart~A. {Sim}, and Henrietta~A. {Hewitt}, \mnras, {\bf
  494}(4), 4914--4929 (June 2020),  {{arXiv:2004.07727}}.

\bibitem{Titchmarsh1924}
E.~C. Titchmarsh, Proceedings of the London Mathematical Society, {\bf
  s2-22}(1), 15--28 (01 1924),
  {{https://academic.oup.com/plms/article-pdf/s2-22/1/15/4372259/s2-22-1-15.pdf}}.

\bibitem{ZHANG2007116}
Xinhong Zhang and Dengke Tong, Applied Mathematics and Computation, {\bf
  193}(1), 116--126 (2007).

\end{thebibliography}

%
%

\end{document}